\documentclass[%
 reprint,
 superscriptaddress,
 amsmath,amssymb,
 aps,
 prapplied,
]{revtex4-2}

\usepackage{graphicx}% Include figure files
\usepackage{dcolumn}% Align table columns on decimal point
\usepackage{bm}% bold math
\usepackage{hyperref}% add hypertext capabilities
\begin{document}

\preprint{APS/123-QED}

\title{Production of multi-oriented polarization for relativistic electron beams \\ via a mutable filter for nonlinear Compton scattering}% Force line breaks with \\

\author{Yuhui Tang}
 \affiliation{State Key Laboratory of Nuclear Physics and Technology, and Key Laboratory of HEDP of the Ministry of Education, CAPT, School of Physics, Peking University, Beijing 100871, China}
\author{Qianyi Ma}
 \affiliation{State Key Laboratory of Nuclear Physics and Technology, and Key Laboratory of HEDP of the Ministry of Education, CAPT, School of Physics, Peking University, Beijing 100871, China}
\author{Jinqing Yu}
 \affiliation{School of Physics and Electronics, Hunan University, Changsha, 410082, China}
\author{Yinren Shou}
 \affiliation{State Key Laboratory of Nuclear Physics and Technology, and Key Laboratory of HEDP of the Ministry of Education, CAPT, School of Physics, Peking University, Beijing 100871, China}
\author{Xuezhi Wu}
 \affiliation{State Key Laboratory of Nuclear Physics and Technology, and Key Laboratory of HEDP of the Ministry of Education, CAPT, School of Physics, Peking University, Beijing 100871, China}
\author{Xueqing Yan}
 \email{x.yan@pku.edu.cn}
 \affiliation{State Key Laboratory of Nuclear Physics and Technology, and Key Laboratory of HEDP of the Ministry of Education, CAPT, School of Physics, Peking University, Beijing 100871, China}
 \affiliation{Beijing Laser Acceleration Innovation Center, Huairou, Beijing, 101400, China}
 \affiliation{CICEO, Shanxi University, Taiyuan, Shanxi 030006, China}
 \affiliation{Institute of Guangdong Laser Plasma Technology, Baiyun, Guangzhou, 510540, China}

\date{\today}% It is always \today, today,

\begin{abstract}
We propose a feasible scenario to directly polarize a relativistic electron beam and obtain overall polarization in various directions through a filter mechanism for single-shot collision between an ultrarelativistic unpolarized electron beam and an ultraintense circularly polarized laser pulse. The electrons are scattered to a large angular range of several degrees and the polarization states of the electrons are connected with their spatial position after the collision. Therefore, we can employ a filter to filter out a part of the scattered electrons based on their position and obtain high-degree overall polarization for the filtered beam. Through spin-considered Monte-Carlo simulations, polarization with a degree up to $62\%$ in arbitrary transverse directions and longitudinal polarization up to $10\%$ are obtained for the filtered beams at currently achievable laser intensity. We theoretically analyze the distribution formation of the scattered electrons and investigate the influence of different initial parameters through simulations to demonstrate the robustness of our scheme. This scenario provides a simple and flexible way to produce relativistic polarized electron beams for various polarization directions.
\end{abstract}

\maketitle

%\tableofcontents
\section{Introduction}
Since the spin properties of relativistic electrons were systematically investigated in the last century~\cite{wigner1939unitary,bargmann1947irreducible,bargmann1948group,shirokov1958group1,shirokov1958group2,shirokov1958group3,shirokov1959relativistic}, extensive applications of spin-polarized electron beams have been developed in the fields of high-energy, nuclear, atomic, and particle physics~\cite{tolhoek1956electron}. The electrons possessing different polarization cause different cross section when interacting with other particles. Taking advantage of this, the polarized relativistic electron beams have played a crucial role in the measurement of the spin structure of neutron~\cite{anthony1993determination} and nucleus~\cite{abe1995precision}, the investigation for parity nonconservation effects~\cite{prescott1978parity,anthony2004observation,jefferson2018precision}, the search for physics beyond the standard model~\cite{moortgat2008polarized}, and the production of polarized photons~\cite{olsen1959photon,martin2012polarization} and highly polarized positrons~\cite{olsen1959photon,abbott2016production} through bremsstrahlung radiation.

Presently, two typical methods are utilized to generate relativistic polarized electron beams in experiments. The first is obtaining transversely polarized electron beams directly from storage rings~\cite{mane2005spin} via the Sokolov-Ternov effect~\cite{sokolov1967synchrotron,bordovitsyn1999synchrotron}. 
Limited by the field strength, this method takes a rather long time, typically several hours, to acquire a high-degree polarization oriented in the opposite direction of the magnetic field~\cite{sun2010polarization}.
If polarization in other directions is required, e.g., longitudinal polarization that more commonly used in practical applications, an additional spin rotator is required to inserted into the beamline to alter the direction of polarization. 
For such ultrarelativistic electron beams with energy above GeV, it is feasible to alter the polarization direction by magnets~\cite{buon1986hera}.
The second method is a multi-stage approach, where nonrelativistic polarized electrons are first extracted from polarized photocathodes~\cite{pierce1975negative,pierce1976photoemission} or spin filters~\cite{batelaan1999optically}, and then accelerated to high energy in linacs. 
The extracted electrons are usually longitudinally polarized, and the Wien filter is a convenient instrument to alter the polarization direction of these low-energy nonrelativistic electrons~\cite{jackson1998classical,steiner2007wien}. In addition, the polarization measurement of relativistic electron beams can be performed utilizing the mechanisms of Mott scattering~\cite{mott1929scattering,gay1992mott}, M{\o}ller scattering~\cite{moller1932theorie,cooper1975polarized,hauger2001high}, Compton scattering~\cite{barber1993hera,beckmann2002longitudinal}, and synchrotron radiation~\cite{belomesthnykh1984observation}.

The above traditional methods for producing relativistic polarized electron beams all require the aid of large-scale accelerator facilities, which leads to a high bar for experimental researches that utilize relativistic polarized electron beams.
With rapid advancement of ultraintense laser technology in recent years, the laser intensity can exceed $10^{23}\,\mathrm{W/cm^2}$~\cite{danson2019petawatt,yoon2019achieving,yoon2021realization}, which makes it possible to produce relativistic polarized electron beams via new methods employing ultraintense laser and get rid of bulky traditional acceleration devices. Therefore, diverse scenarios have been prospectively proposed under the theoretical and simulating studies. For instance, when an initially unpolarized ultrarelativistic electron beam head-on collides with an ultraintense elliptically polarized laser pulse, the electron beam is split into two oppositely transversely polarized parts with a splitting angle of about tens of milliradians due to nonlinear Compton scattering~\cite{li2019ultrarelativistic}. 
Taking advantage of the asymmetry in the field of a two-color or few-cycle ultraintense laser pulse, an unpolarized ultrarelativistic electron beam can acquire overall transverse polarization from the collision with the laser pulse~\cite{song2019spin,seipt2018theory,seipt2019ultrafast}.
Besides, there are also schemes of producing polarized electron beams via hydrogen halide targets prepolarized by ultraviolet photodissociation~\cite{rakitzis2003spin,rakitzis2004pulsed,sofikitis2008laser,sofikitis2018ultrahigh} in laser-driven or particle-driven plasma wakefield acceleration~\cite{wen2019polarized,wu2019polarized,wu2019polarized1}, and a filter can be deployed for wakefield acceleration to increase the polarization degree of electron beams~\cite{wu2020spin}.
%尾场加速会受到电量限制……

\begin{figure}
\includegraphics[keepaspectratio=true,width=86mm]{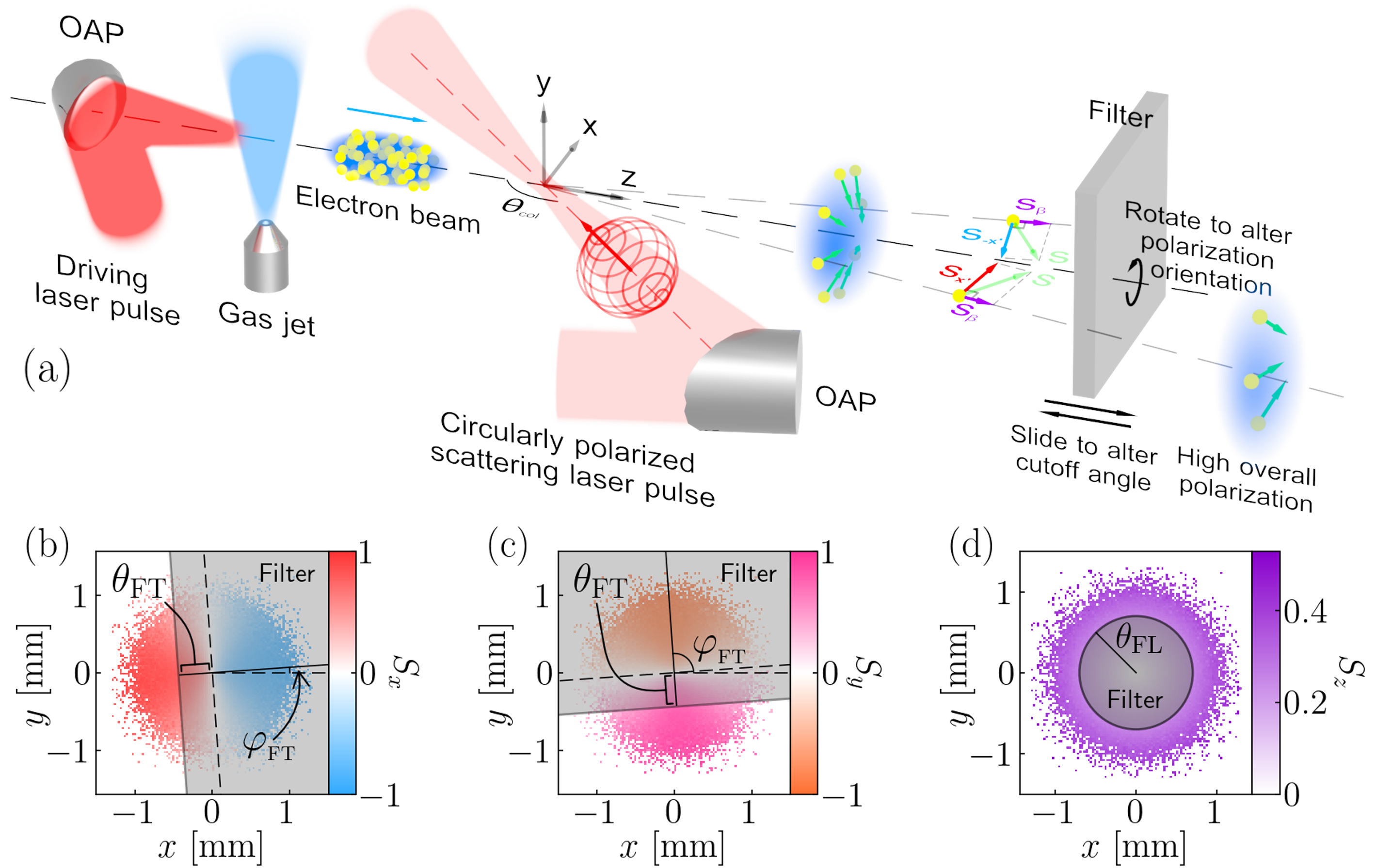}
\caption{\label{fig:1} Scenario of producing relativistic electron beams with multi-oriented polarization. (a) Schematic layout.
(b)-(d) The filter settings to obtain high degree of (b) $S_x$, (c) $S_y$, and (d) $S_z$, where the spatial distribution of the electrons is at $10\ \mathrm{ps}$ after the collision ($3\ \mathrm{cm}$ behind the collision point). $\theta_\mathrm{FT}$ and $\varphi_\mathrm{FT}$ are the cutoff scattering angle and the orientation of the filter for transverse polarization, respectively. $\theta_\mathrm{FL}$ is the cutoff scattering angle of the filter for longitudinal polarization.}
\end{figure}

In this article, we propose a filter mechanism for the collision between an initially unpolarized electron beam and an ultraintense circularly polarized laser pulse to obtain relativistic polarized electron beams with various polarization directions. By filtering out a part of scattered electrons based on their position after the collision, we can easily obtain overall polarization in any transverse direction and longitudinal direction without any additional spin rotator, as shown in Fig.~\ref{fig:1}. The entire processes of the generation, polarization, and utilization of relativistic electron beams can be accomplished within tens of centimeters, which is much compact than the methods resorting to traditional acceleration devices. Therefore, this new method of producing relativistic polarized electron beams may be a better choice for some applications in high-energy physics.
Through spin-considered Monte-Carlo simulations, we obtain overall polarization with a degree up to $62\%$ in arbitrary transverse directions and longitudinal polarization around $10\%$ at currently achievable laser intensities.
For the electrons of several hundred MeV with a broad energy spectrum after nonlinear Compton scattering, the traditional spin rotation system, no matter Wien filters or magnets, cannot be applied to such an electron beam directly, so the scheme with the ability to produce multi-oriented polarization shows superiority in the circumstances.
%对于百MeV左右的电子，使用磁铁的旋转方案难以直接实施，而韦恩过滤器方案对于如此高能的电子，其装置制造难度会大大增加，并且他并不能应用于散射后产生的宽谱电子。
Furthermore, we also investigate the simulation results under different initial parameters and demonstrate the robustness of our scheme.

\section{Simulation methods and setups \label{sec:setups}}
The scenario is illustrated in Fig~\ref{fig:1}. A collimated relativistic electron beam propagates along the $z$-axis and collides with an ultraintense circularly polarized laser beam at the origin. The initial electron beam can be extracted from laser-driven plasma wakefield acceleration, where electron beams with energies up to $8\,\mathrm{GeV}$ are obtained with current state-of-the-art experimental techniques~\cite{leemans2014multi,gonsalves2019petawatt}.
The scattering laser propagating in the horizontal plane ($xz$-plane) is focused at the origin, and the angle between the laser propagation direction and the electron beam propagation direction is $\theta_\mathrm{col}$.
Since the filter need to be placed on the route of the electron beam, a collision angle that does not affect post-processing of scattered electrons is chosen rather than head-on collision.

When electrons interact with an ultraintense laser beam whose normalized laser field $a_0 \gg 1$, nonlinear Compton scattering is dominant in the interaction~\cite{gol1964intensity,nikishov1964quantum,ritus1985quantum}, where $a_0 = |e|E_0/m_e \omega_L c$, $e$ is the electron charge, $m_e$ is the rest mass of electron, $c$ is the speed of light, $E_0$ and $\omega_L$ are the amplitude and frequency of laser field, respectively. 
In the interaction, it is easy to reach the condition of quantum regime $\chi_e \gtrsim 1$, where $\chi_e = |e|\hbar\sqrt{(F_{\mu\nu}p^\nu)^2}/m_e^3 c^4$ is the invariant quantum efficiency parameter, $\hbar$ is the reduced Planck constant, $F_{\mu\nu}$ is the electromagnetic field tensor, and $p^\nu$ is the electron four-momentum. The stochastic nature of high-energy photon emission becomes important in the quantum regime, so a Monte-Carlo method is more appropriate for simulations~\cite{gong2021retrieving}. Therefore, to investigate the spin polarization distribution of the scattered electrons, we develop a polarization-vector-based Monte-Carlo code~\cite{gong2019radiation,tang2021radiative}, where both nonlinear Compton scattering and classical spin precession (i.e., Thomas-Bargmann-Michel-Telegdi equation~\cite{thomas1926motion,thomas1927kinematics,bargmann1959precession}) are considered.

In our code, the electron spin is described as a three-dimensional normalized polarization vector $\mathbf{S}$ in the rest frame of the electron. The projection of $\mathbf{S}$ indicates the expected value when measuring polarization along the projection axis. $|\mathbf{S}| = 1$ stands for a fully polarized pure state, $0 < |\mathbf{S}| < 1$ stands for a partially polarized mixed state, and $|\mathbf{S}| = 0$ stands for a completely unpolarized state. 

Applying local constant field approximation, the differential photon emission rate of nonlinear Compton scattering can be derived as~\cite{seipt2018theory}
\begin{equation}
\begin{aligned}
    \frac{\mathrm{d}^2 N_\gamma}{\mathrm{d}\chi_\gamma\mathrm{d}t} =\ & \frac{\alpha}{\sqrt{3} \pi\tau_c \gamma \chi_e} \bigg [ \left (2+3\chi_\gamma y \right ) K_{\frac{2}{3}}(2y) \\
    &- \mathrm{Int}K(2y) - \frac{\chi_\gamma}{\chi_e} S_b K_{\frac{1}{3}}(2y) \bigg ],
\end{aligned}
\end{equation}
where $\tau_c = \hbar/m_e c^2$ is the Compton time, $\alpha= e^2/\hbar c \approx 1/137$ is the fine-structure constant, $\gamma = 1/\sqrt{1-\boldsymbol{\beta}^2}$ is the Lorentz factor, $\boldsymbol{\beta} = \mathbf{v}/c$ is the velocity of electron, $y = \chi_\gamma / [3 \chi_e (\chi_e - \chi_\gamma)]$, $K_{\nu}(x)$ is the modified Bessel functions of the second kind, $\mathrm{Int}K(x) = \int_{x}^\infty K_{\frac{1}{3}}(z)\mathrm{d}z$, and $\chi_\gamma$ is a quantum parameter related to the photon energy. $S_b = \mathbf{S} \cdot \hat{\mathbf{b}}$, where $\hat{\mathbf{b}} = \mathbf{B}_\mathrm{RF}/|\mathbf{B}_\mathrm{RF}|$ is the direction of magnetic field in the rest frame, and the field in the rest frame can be obtained through the Lorentz transformation from the field in the laboratory frame.

In simulations, the electron particles are moved by the Lorentz force in preset electromagnetic fields and emit photons as following. First, the instant of the emission is decided by the optical depth $\tau$ and the final optical depth $\tau_f$ assigned to each particle. At the beginning of simulations, $\tau$ is set to $0$, $\tau_f$ is sampled from $\tau_f = -\ln \xi_1$, where $\xi_1$ is a uniform random number in $(0,1)$. The optical depth evolves according to ${\mathrm{d}\tau}/{\mathrm{d}t} = \int_0^{\chi_e}{ ({\mathrm{d}^2N_\gamma}/{\mathrm{d}\chi_\gamma \mathrm{d}t}) \mathrm{d}\chi_\gamma }$~\cite{duclous2010monte}. Every time $\tau$ reaches $\tau_f$, $\tau$ is reset to $0$, $\tau_f$ is resampled by the same way, and a photon is emitted along $\hat{\boldsymbol{\beta}}$. Next, the $\chi_\gamma$ of the photon is decided by another uniform random number $\xi_2 \in (0,1]$ from $
\xi_2 = [{\int_0^{\chi_\gamma}{ ({\mathrm{d}^2N_\gamma}/{\mathrm{d}\chi_\gamma \mathrm{d}t})  \mathrm{d}\chi_\gamma }}]/[{\int_0^{\chi_e}{ ({\mathrm{d}^2N_\gamma}/{\mathrm{d}\chi_\gamma \mathrm{d}t})  \mathrm{d}\chi_\gamma }}]
$. Then, the photon energy can be calculated by $\hbar\omega_\gamma = \gamma m_e c^2 \chi_\gamma/\chi_e$. According to momentum conservation, the electron momentum becomes $\mathbf{p}' = \mathbf{p}-(\hbar\omega_\gamma/c)\mathbf{\hat p}$ after the emission. In addition, the electron polarization $\mathbf{S}$ is subject to a step alteration after the emission, i.e.
\begin{equation}
\begin{aligned}
    \Delta \mathbf{S} =& \bigg \{ \left[ \frac{\chi_\gamma}{\chi_e} S_b K_{\frac{1}{3}}(2y) - 3\chi_\gamma y \mathrm{Int}K(2y) \right] S_\beta \hat{\boldsymbol{\beta}} \\
    &+ \left[ \frac{\chi_\gamma}{\chi_e} S_b K_{\frac{1}{3}}(2y) - 3\chi_\gamma y K_{\frac{2}{3}}(2y) \right] S_e \hat{\mathbf{e}} \\
    &+ \left[\left( \frac{\chi_\gamma}{\chi_e}S_b^2 - 3\chi_e y \right) K_{\frac{1}{3}}(2y) - 3\chi_\gamma y S_b K_{\frac{2}{3}}(2y) \right] \hat{\mathbf{b}} \bigg \} \\
    &/ \left [ \left (2+3\chi_\gamma y \right ) K_{\frac{2}{3}}(2y) - \mathrm{Int}K(2y) - \frac{\chi_\gamma}{\chi_e} S_b K_{\frac{1}{3}}(2y) \right ],
\end{aligned}\label{eq:delta_S}
\end{equation}
where $\hat{\mathbf{e}} = \mathbf{E}_\mathrm{RF}/|\mathbf{E}_\mathrm{RF}|$ is the direction of electric field in the rest frame.

Besides, the polarization evolves under the combined effect of the classical spin precession and the no-photon-emission part of radiation effect at all times~\cite{tang2021radiative}, i.e.
\begin{eqnarray}
    \frac{\mathrm{d} \mathbf{S}}{\mathrm{d} t} &=& \boldsymbol{\Omega} \times \mathbf{S} + \frac{\alpha \left( \hat{\mathbf{b}} - S_b \mathbf{S} \right)}{\sqrt{3} \pi\tau_c \gamma \chi_e} \int_0^{\chi_e} \frac{\chi_\gamma}{\chi_e} K_{1/3}(2y) \mathrm{d}\chi_\gamma, \\
    \boldsymbol{\Omega} &=&- \frac{e}{m_e c} \Bigg [ \left( a(\chi_e) + \frac{1}{\gamma} \right) \mathbf{B} - \frac{a(\chi_e) \gamma}{\gamma+1} \left( \boldsymbol{\beta} \cdot \mathbf{B} \right) \boldsymbol{\beta} \nonumber\\
    & &– \left( a(\chi_e) + \frac{1}{\gamma+1} \right) \boldsymbol{\beta} \times \mathbf{E} \Bigg ], \nonumber
\end{eqnarray}
where $a(\chi_e) = ({\alpha}/{\pi\chi_e})\int^\infty_0 [{u}/{(1+u)^3}]L_{\frac{1}{3}}( {2u}/{3\chi_e} )\mathrm{d}u$ is the anomalous magnetic moment of the electron taking the radiative correction into consideration~\cite{bauier1972radiative}, $L_{\frac{1}{3}}(z) = \int^\infty_0 \sin\left[ ({3z}/{2}) \left(x+{x^3}/{3}\right) \right] \mathrm{d}x$.

The simulation method is valid under the assumptions that $\gamma \gg 1$, $a_0 \gg 1$, and the electromagnetic fields are small with respect to the Schwinger limit field $E_\mathrm{crit} = m_e^2 c^3/e \hbar \approx 1.32\times 10^{18}\,\mathrm{V/m}$~\cite{seipt2018theory}. Our code is verified by reproducing the result of Sokolov-Ternov effect and other spin polarization models~\cite{bordovitsyn1999synchrotron,guo2020stochasticity,li2020production}.

In the simulation, the initial electron beam possesses a width of $1\,\mathrm{\mu m}$, a length of $5\,\mathrm{\mu m}$, an angular divergence of $3\,\mathrm{mrad}$, and an energy distribution centered around $5\,\mathrm{GeV}$ with a spread of $5\%$, while the circularly polarized scattering laser beam possesses a wavelength of $800\,\mathrm{nm}$, a Gaussian envelope with a waist radius of $5\,\mathrm{\mu m}$, an intensity peaking at $2.14\times10^{22}\,\mathrm{W/cm^2}$ ($a = a_0/\sqrt{2} = 100/\sqrt{2}$), and a hyperbolic secant temporal profile with duration of $\tau_\mathrm{FWHM} = 16\,\mathrm{fs}$. The collision angle is set as $\theta_{\mathrm{col}} = 135^{\circ}$ and 10 million particles are simulated in total. The quantum efficiency parameter $\chi_e$ can be derived as $\chi_e \simeq \gamma (1 - \cos \theta_\mathrm{col}) E_0 / E_\mathrm{crit}$, so the maximum $\chi_e \approx 3.6$. 

\section{Results and discussions}

\begin{figure}
\includegraphics[keepaspectratio=true,width=86mm]{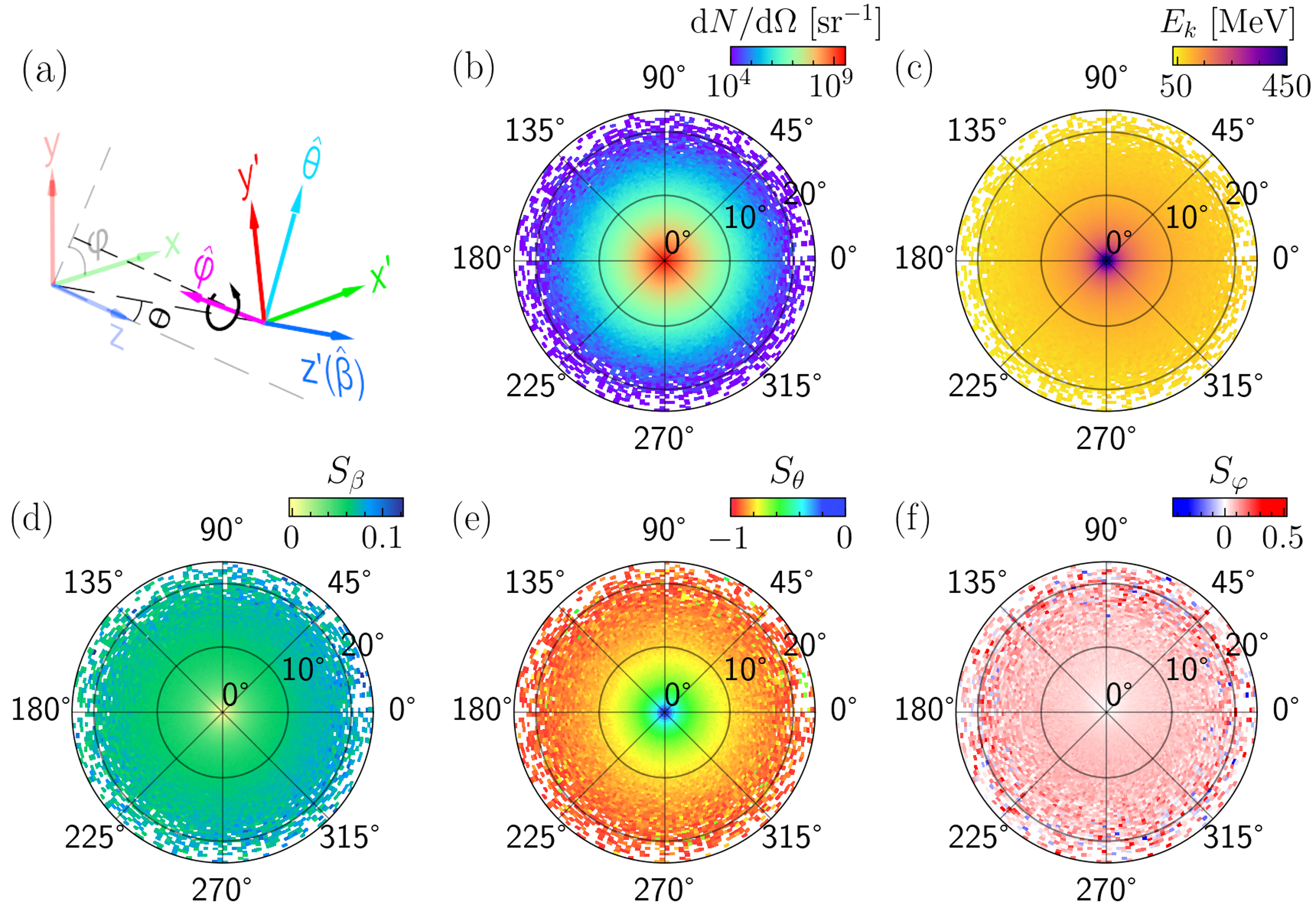}
\caption{\label{fig:2} (a) The relation of the coordinates and the vectors. (b) The number per solid angle $\mathrm{d}N/\mathrm{d}\Omega$, (c) the kinetic energy $E_k$, (d) the longitudinal polarization $S_\beta$, (e) the radial polarization $S_\theta$, and (f) the azimuthal polarization $S_\varphi$ versus the scattering direction of the electrons.}
\end{figure}

\subsection{Filter settings}
After the interaction, the electrons are scattered to a large angular range of tens of degrees and form an axisymmetric distribution with respect to the $z$-axis (i.e., the original direction of beam motion), as the simulation results shown in Fig.~\ref{fig:2}. The larger the scattering angle, the smaller the number of electrons, and the lower the electron energy. Most of the electrons are in the energy range of tens to hundreds of MeV and still keep relativistic. For electrons in this energy range, when they are subjected to an electromagnetic field and change their motion directions, their spins are usually synchronously rotated. Considering the scattering directions are distributed over a large angular range, the statistics of average polarization of all the electrons in the laboratory frame $(x, y, z)$ is pointless. Therefore, we utilize a modified frame $(x', y', z')$ for each particle to analyze the spin polarization, as shown in Fig.~\ref{fig:2}(a). The modified frame $(x', y’, z’)$ can be obtained by rotating the frame $(x, y, z)$ an angle $\theta$ around $\hat{\varphi}$, where $\theta$ is the angle between the motion direction $\hat{\boldsymbol{\beta}}$ and $z$-axis, $\hat{\varphi}$ is the direction perpendicular to $\hat{\boldsymbol{\beta}}$ and $z$-axis. The average polarization calculated in the frame $(x', y’, z’)$ is equivalent to the corresponding polarization in the frame $(x, y, z)$ when all the electrons are focused and move parallel to $z$-axis. Through the polarization distribution we can see that the polarization is mainly radial. The electrons with scattering angles above $10^\circ$ mostly possess a radial polarization of around $80\%$ degree, while the longitudinal polarization and azimuthal polarization are about $7\%$ and $10\%$ respectively. However, the overall polarization of all scattered electrons is less than $3\%$ for longitudinal polarization and almost nil for transverse polarization because of the axial symmetry of the polarization distribution.

Fortunately, the large angular distribution allows us to easily select the polarized electrons with the same polarization direction, and then obtain an electron beam with high-degree overall polarization. Since the electron beam is collimated before the collision, the angular divergence and spatial size of the initial beam have little effect on the final distributions at such a large scattering angular divergence. When the electron beam travels a distance after the collision, the symmetry of the distributions is transferred to the spatial position of the scattered electrons, as shown in Fig.~\ref{fig:1}(b)-(d).
Then, we can set a filter for the required polarization direction to filter out electrons that possess polarization in the opposite direction or polarization with low degree, so as to obtain an electron beam with high polarization in the required direction. For transverse polarization, a rectangular filter can be employed to filter out the scattered electrons on one side, where the orientation of the filter is $\varphi_\mathrm{FT}$ and the cutoff scattering angle is $\theta_\mathrm{FT}$. Due to the axisymmetric nature of the scattering distribution, the overall polarization in other transverse directions can be easily obtained by rotating the filter around the $z$-axis. For longitudinal polarization, a circular filter can be employed to filter out electrons of lower polarization in the central area, where the cutoff scattering angle is $\theta_\mathrm{FL}$. The cutoff scattering angles for both filters can be precisely controlled by sliding the filters along the $z$-axis.

\begin{figure}
\includegraphics[keepaspectratio=true,width=86mm]{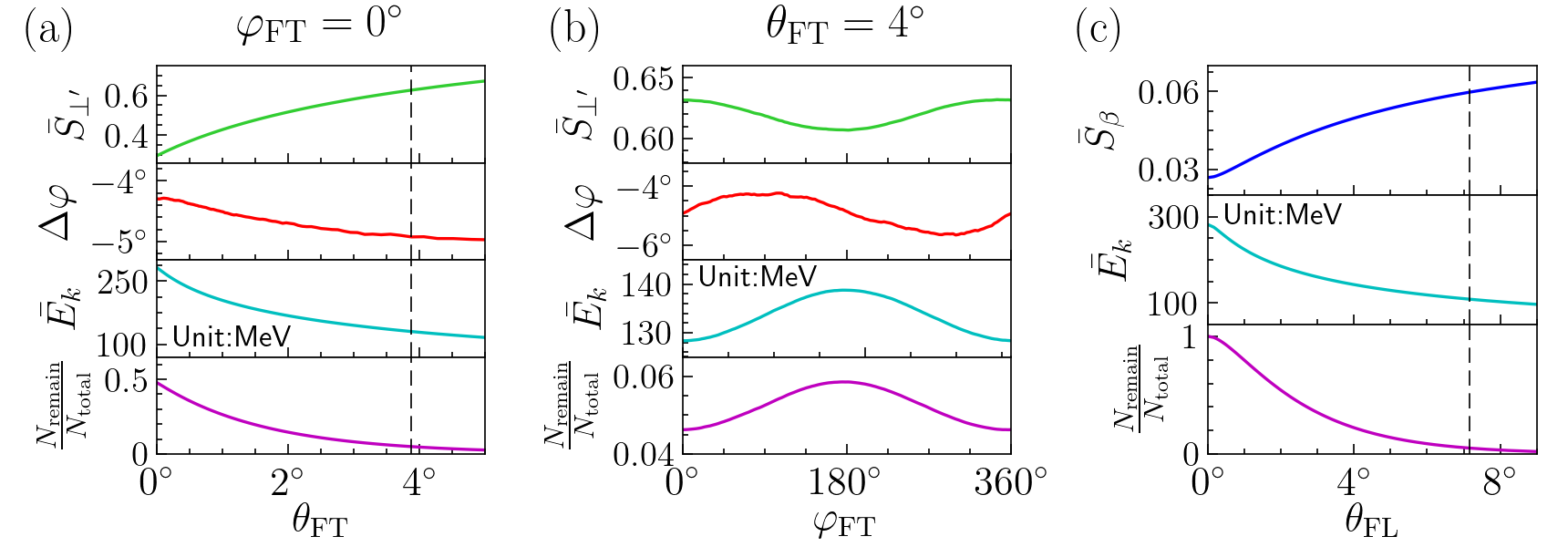}
\caption{\label{fig:2c} The results of the filtered beams after applying (a) the filter for transverse polarization with $\varphi_\mathrm{FT} = 0^\circ$ and different $\theta_\mathrm{FT}$, (b) the filter for transverse polarization with $\theta_\mathrm{FT}=4^\circ$ and different $\varphi_\mathrm{FT}$, and (c) the filter for longitudinal polarization with different $\theta_\mathrm{FL}$. $\Delta \varphi$ is the angular difference between the direction of the overall transverse polarization $\mathbf{S}_{\perp'}$ and the filter orientation $\varphi_\mathrm{FT}$. The values corresponding to $N_\mathrm{remain}/N_\mathrm{total} = 5\%$ are marked with the dash lines.}
\end{figure}

The results of the filtered beam at different cutoff scattering angles and filter orientations are shown in Fig.~\ref{fig:2c}. The larger the cutoff scattering angle, the higher the overall polarization degree of the filtered beam, but meanwhile the smaller the electron number and the lower the average energy of the filtered beam. For transverse polarization, about $62\%$ polarization can be obtained with the electron number ratio $N_\mathrm{remain}/N_\mathrm{total} = 5\%$. In this condition, the cutoff scattering angle is about $4^\circ$ and the average energy is about $130\,\mathrm{MeV}$. Because of the azimuthal polarization emerging under the effect of spin precession, the direction of the overall transverse polarization is not completely parallel to the filter direction $\varphi_\mathrm{FT}$, but has an angular difference of about $5^\circ$, which needs to be taken into account to obtain the optimal results.
Since the scattering laser and the electron beam do not head-on collide along the same axis, the results of different filter orientations are slightly different with the same cutoff scattering angle, but this does not affect our conclusions. For longitudinal polarization, about $6\%$ polarization degree can be obtained with $5\%$ of particles remaining, which is more than twice that of the case without filtering. In this case, the cutoff scattering angle is about $7^\circ$ and the average energy is about $108\,\mathrm{MeV}$.

\subsection{Theoretical analyses\label{sec:analyses}}
The formation of the distributions can be explained by the electron motion in a plane electromagnetic wave. The electron is assumed to propagate along $z$-axis initially and possess a momentum $p_0$, while the plane wave is assumed to propagate parallel to the $xz$-plane with the collision angle $\theta_\mathrm{col}$. The electron motion is easy to be solved in the wave frame $(x_L, y_L, z_L)$ where the wave propagates along $z_L$-axis. The frame $(x_L, y_L, z_L)$ can be obtained by rotating the frame $(x,y,z)$ around $y$-axis by $\theta_\mathrm{col}$.
%\begin{equation}
%\begin{bmatrix}x_L\\y_L\\z_L\end{bmatrix} = 
%\begin{bmatrix}
%\cos \theta_\mathrm{col} & 0 & -\sin \theta_\mathrm{col}\\
%0 & 1 & 0\\
%\sin \theta_\mathrm{col} & 0 & \cos \theta_\mathrm{col}
%\end{bmatrix}
%\begin{bmatrix}x\\y\\z\end{bmatrix}.\label{eq:rotation}
%\end{equation}
The wave can be represented by the normalized magnetic vector potential $\mathbf{a} = |e|\mathbf{A}/m_e c^2 = \delta a_0 \cos \phi \hat{\mathbf{x}}_L + \sqrt{1-\delta^2} a_0 \sin \phi \hat{\mathbf{y}}_L$, where $\phi = \omega_L t - \mathbf{k}\cdot\mathbf{r} + \phi_0$ is the phase of the wave, and $\delta$ is a polarization parameter such that $\delta = \pm 1/\sqrt{2}$ for a circularly polarized wave. Then, we can obtain the expression for the momenta in the frame $(x_L, y_L, z_L)$ as~\cite{gibbon2005short}
\begin{equation}
\begin{aligned}
    p_{x_L} =&\ \mathbf{a} \cdot \hat{\mathbf{x}}_L - p_0 \sin \theta_\mathrm{col}, \\
    p_{y_L} =&\ \mathbf{a} \cdot \hat{\mathbf{y}}_L, \\
    p_{z_L} =&\ p_0 \cos \theta_\mathrm{col} + \frac{\mathbf{a}^2 - 2p_0 \sin \theta_\mathrm{col} \mathbf{a} \cdot \hat{\mathbf{x}}_L}{2(\gamma_0 - p_0 \cos \theta_\mathrm{col})},
\end{aligned}
\end{equation}
where $\gamma_0 = \sqrt{1+p_0^2}$ is the initial Lorentz factor of the electron. Note that here we use the dimensionless momentum (i.e. $p \to p/m_e c$) to simplify the expression. For an initially ultrarelativistic electron (i.e. $\gamma_0 \gg 1$), we can take the approximation of $\gamma_0 \approx p_0$ and utilize coordinate transformation to obtain the expression in the frame $(x,y,z)$, i.e.
\begin{eqnarray}
    p_x &=& - \delta a_0 \sin\phi' + \frac{a_0^2}{2\gamma_0 \tan(\theta_\mathrm{col}/2)}, \nonumber \\
    p_y &=& \sqrt{1-\delta^2} a_0 \cos\phi', \label{eq:p}\\
    p_z &=& p_0 + \frac{a_0^2}{2\gamma_0(1/\cos\theta_\mathrm{col} - 1)} -  \frac{\delta a_0}{\tan(\theta_\mathrm{col}/2)} \sin\phi', \nonumber
\end{eqnarray}
where $\phi' \approx (1 - \cos \theta_\mathrm{col}) \omega_L t + \phi'_0$.

\begin{figure}
\includegraphics[keepaspectratio=true,width=86mm]{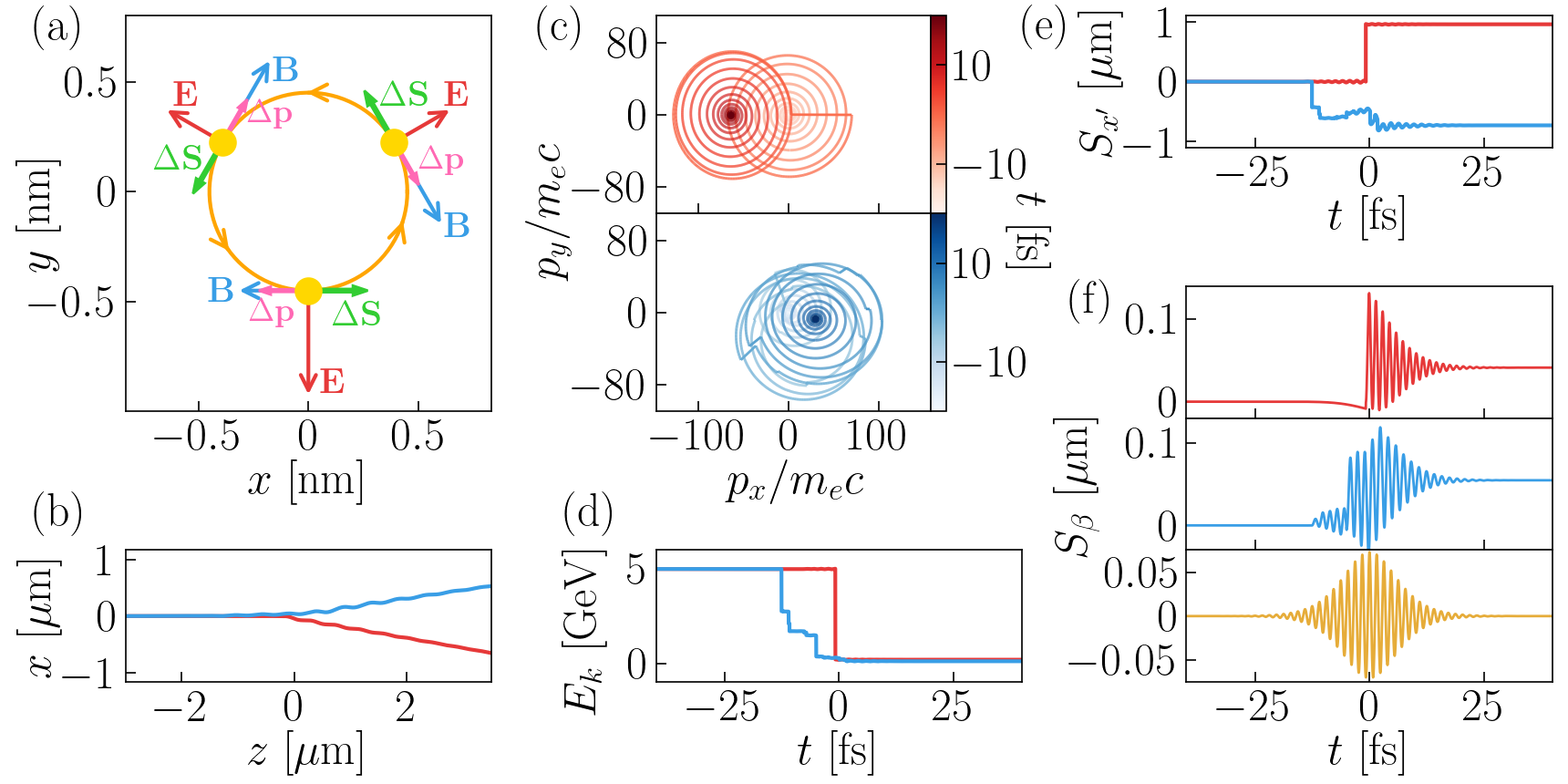}
\caption{\label{fig:3} (a) Illustrative diagram of the electron motion and radiation. (b)-(f) The evolution of an explanatory particle that only emits one high-energy photon during the interaction (the red lines), a selected particle in the Monte-Carlo simulation (the blue lines), and a fully transversely polarized electron that is affected by the spin precession but not the radiation effect (the yellow line).}
\end{figure}

For $\gamma_0 \gg a_0$ and $\theta_\mathrm{col} \in [90^\circ, 180^\circ]$, the second term of $p_x$ is far less than the amplitude of the periodic term, so the motion in the $xy$-plane can be sketchily drawn as a circle for a circularly polarized wave, as shown in Fig.~\ref{fig:3}(a). Since the electron circles synchronously with the wave, it can be calculated that the direction of the magnetic field in the rest frame, i.e. $\mathbf{B}_{\mathrm{RF}} = \gamma \left[ \mathbf{B} - \boldsymbol{\beta}\times\mathbf{E} - \frac{\gamma}{\gamma+1}\left(\boldsymbol{\beta} \cdot \mathbf{B} \right) \boldsymbol{\beta} \right]$, is always close to the opposite direction (for $\delta=1/\sqrt{2}$) of the transverse momentum. According to Eq.~\eqref{eq:delta_S}, when the electron emits a high-energy photon, the electron is significantly polarized towards $-\hat{\mathbf{b}}$ and depolarized along the directions of $\hat{\boldsymbol{\beta}}$ and $\hat{\mathbf{e}}$. The higher the photon energy, the larger the polarization alteration. Thus, whenever an unpolarized electron emits a high-energy photon, it obtains a high-degree polarization in the same direction as the current transverse momentum. On the other hand, the electron momentum obtains a step alteration $\Delta \mathbf{p}$ in the opposite direction of current momentum due to the radiation reaction. The $\Delta \mathbf{p}$ is beyond the oscillating momentum, which changes the neutral direction and makes the electron be scattered out of the original direction after the interaction. Therefore, if the electron emits only one and high-energy photon in the whole interaction, as the explanatory particle shown by red lines in Fig.~\ref{fig:3}, the electron will eventually obtain a high-degree polarization towards $-\hat{\boldsymbol{\theta}}$. In actual simulations, the electron always emits multiple photons throughout the interaction. However, for a short laser pulse, the ultimate state of the electron is mostly decided by the emission of a few highest-energy photons, as the simulated particle shown by blue lines in Fig.~\ref{fig:3}, so the qualitative discussion above is still applicable.

As for longitudinal polarization, generally, the electron cannot obtain the growth of longitudinal polarization from a rapidly changing periodic field such as a laser field with the sole presence of the spin precession. The evolution of longitudinal polarization under the spin precession effect can be written as~\cite{jackson1998classical}
\begin{equation}
    \frac{\mathrm{d}S_\beta}{\mathrm{d}t} = - \frac{e}{m_e c} \mathbf{S}_\perp \cdot \left[ a(\chi_e) \left( \hat{\boldsymbol{\beta}} \times \mathbf{B} + \beta \mathbf{E} \right) + \left(\beta - \frac{1}{\beta}\right) \mathbf{E} \right], \label{dS_beta}
\end{equation}
where $\mathbf{S}_\perp$ is the transverse component of $\mathbf{S}$. For the interaction between a relativistic electron and a circularly polarized laser field, when the polarization is mainly transverse (i.e., $S_\perp \gg S_\beta$), $S_\beta$ can be approximately solved as
\begin{equation}
    S_\beta = \frac{S_\perp a(\chi_e) a_0}{\sqrt{2}}\cos\phi'+S_{\beta 0},
\end{equation}
where $S_{\beta 0}$ is the initial longitudinal polarization of the electron. It can be seen that the longitudinal polarization oscillates synchronously with the laser field and returns to the initial value $S_{\beta 0}$ when the electron leaves the laser field, as shown by the yellow line in Fig.~\ref{fig:3}(f). However, taking the radiation effect into account, the results will be different. When a high-energy photon is emitted, the polarization is generated in the direction of $-\hat{\mathbf{b}}$, so we can calculate that ${\mathrm{d}S_\beta}/{\mathrm{d}t} \approx 0$ according to the relationship of the vector directions, which implies that the oscillation of $S_\beta$ is at the peak or trough. In other words, the neutral position of the oscillation of longitudinal polarization is altered after the emission. When the electron moves away from the laser pulse, the longitudinal polarization has increased compared to the beginning, as shown by the red line in Fig.~\ref{fig:3}(f). Because of the symmetry, the neutral position moves towards the same direction whenever the emission occurs. Therefore, the effect accumulates for multiple emission, as shown by the blue line in Fig.~\ref{fig:3}(f).

\subsection{Influence of different initial conditions}
\begin{figure}
\includegraphics[keepaspectratio=true,width=86mm]{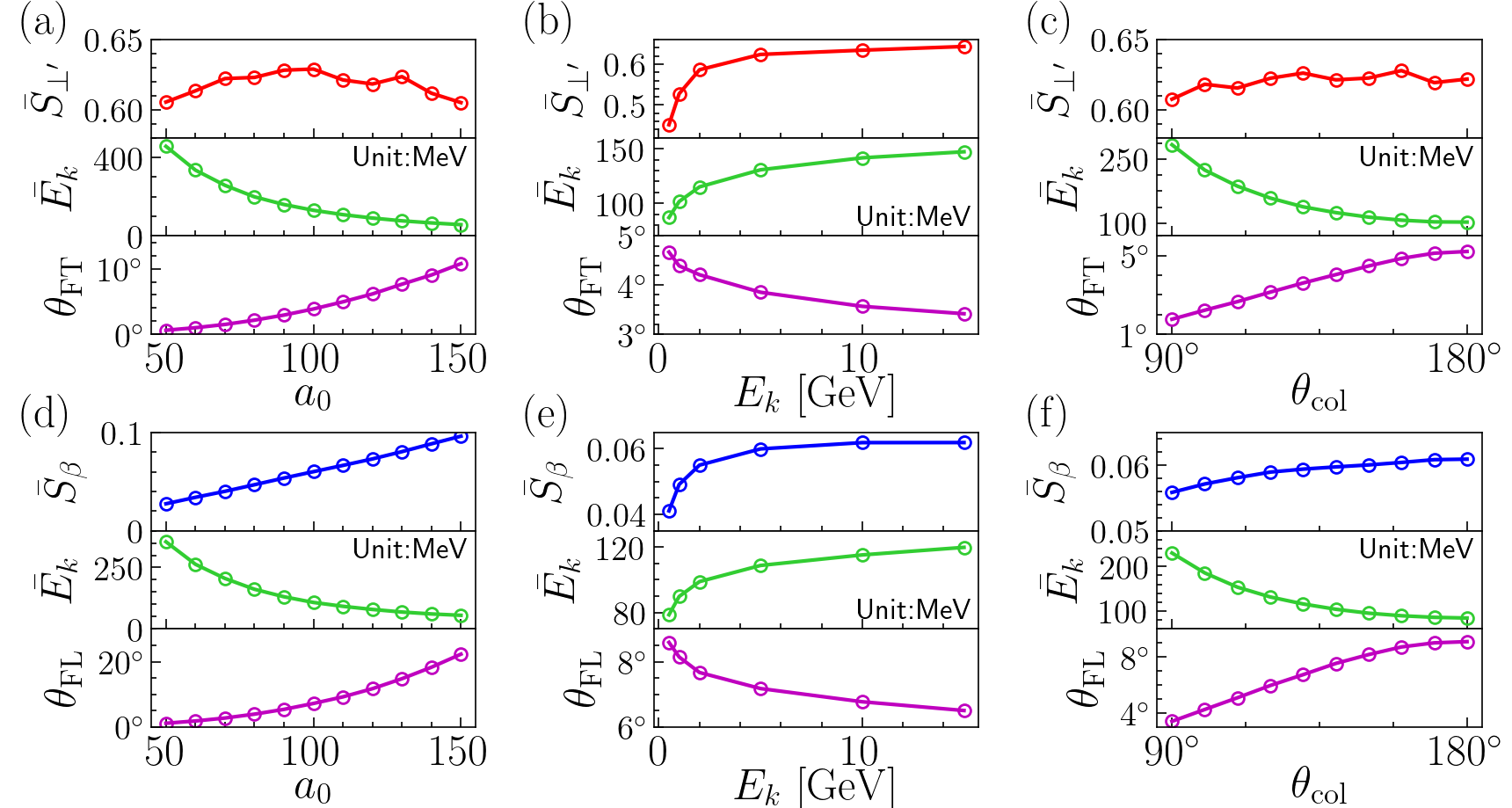}
\caption{\label{fig:4} The simulation results of the filtered beam with $5\%$ of particles remaining under different laser intensity $a_0$, initial beam energy $E_k$, and collision angle $\theta_\mathrm{col}$. (a)-(c) The filter for transverse polarization with $\varphi_\mathrm{FT} = 0^\circ$. (d)-(f) The filter for longitudinal polarization.
}
\end{figure}

In order to test the robustness of our scheme, we investigate the effect of different initial parameters, and the results of the filter beam with $5\%$ of particles remaining are shown in Fig.~\ref{fig:4}. Except for the parameters in the figure, all other initial parameters are the same as those described in Sec.~\ref{sec:setups}. The electron has a high possibility to emit photons with energy comparable to the electron itself and make itself close to the polarization limit once the quantum regime $\chi_e \sim 1$ is reached~\cite{tang2021radiative}. A further increase in $\chi_e$ does not make much difference in polarization. Therefore, transverse polarization with a degree above $60\%$ is obtained under all laser intensities $a_0$ from $50$ to $150$. According to the analyses in Sec.~\ref{sec:analyses}, the longitudinal polarization is transferred from the transverse polarization, and the final $S_\beta$ is proportional to the laser intensity, which is also confirmed by our results. We obtain around $10\%$ longitudinal polarization at $a_0 = 150$ (i.e., $I_0 = 6.42 \times 10^{22}\,\mathrm{W/cm^2}$, which can be achieved by current laser technologies). However, it should be noted that the larger the laser intensity, the stronger the radiation in the laser field, and the larger the radiation reaction force subjected to the electron. If the laser intensity is too large, the electrons will be scattered to every direction and lose almost all of their energy, which makes them unable to be utilized as a beam. Therefore, it is impracticable to blindly increase laser intensity to obtain higher longitudinal polarization.
As for initial beam energy $E_k$, when the beam energy reach about $5\,\mathrm{GeV}$, the further increase in initial energy does not make a significant increase in polarization, which is also because the quantum regime $\chi_e \sim 1$ has been reached. The increase in initial energy can not remarkably increase the energy of final beam either, but it also implies that the method is efficient for the beam with a wide energy spread. Moreover, the average energy of the final beam is higher at a smaller collision angle $\theta_\mathrm{col}$, but the degree of polarization hardly decreases. Therefore, decreasing collision angle may be a convenient way to increase the energy of final beams, as long as the transverse motion range of the electrons is smaller than the spatial size of the laser and the quantum regime $\chi_e \gtrsim 1$ is reached. From these simulation results, we can see that our scheme can be effectively applied for a variety of different initial parameters, thus demonstrating the robustness of our filtering method.

\section{Conclusions}
In conclusion, we propose a filter mechanism for the collision between an initially unpolarized ultrarelativistic electron beam and an ultraintense circularly polarized laser pulse. After the collision, the polarization of the electrons is connected with the motion direction and spatial position, thus we can filter out a part of the scattered electrons based on their spatial position to obtain a relativistic polarized electron beam with high-degree polarization in a certain direction. The entire production process of relativistic polarized electron beams can be realized within tens of centimeters in our scheme, which is much compact compared with the traditional methods resorting to large-scale accelerator devices. Moreover, polarization in various directions can be obtained in the same setup without the help of additional spin rotation devices. Through numerical simulation utilizing the self-developed spin-considered Monte Carlo code, we obtain up to $62\%$ polarization in any transverse direction and about $10\%$ polarization in longitudinal direction for the filtered beam at currently achievable laser intensities. We reveal the reason for the formation of the polarization distribution by analyzing the motion and spin evolution of relativistic electrons in a plane electromagnetic wave. Finally, the influence of different initial parameters on the results is investigated to demonstrate the robustness of our method. This new method for producing relativistic polarized electron beams is practicable to be implemented experimentally and has potential applications in high-energy physics.

\begin{acknowledgments}
The work has been supported by the NSFC (Grants No. 11921006 and No. 11535001) and the National Grand Instrument Project (Grant No. 2019YFF01014400). J.Y. has been supported by the Natural Science Foundation of Hunan Province (Grant No. 12175058). Simulations were supported by the High-Performance Computing Platform of Peking University.
\end{acknowledgments}

%\nocite{*}

\bibliography{apssamp}% Produces the bibliography via BibTeX.

\begin{thebibliography}{66}%
\makeatletter
\providecommand \@ifxundefined [1]{%
 \@ifx{#1\undefined}
}%
\providecommand \@ifnum [1]{%
 \ifnum #1\expandafter \@firstoftwo
 \else \expandafter \@secondoftwo
 \fi
}%
\providecommand \@ifx [1]{%
 \ifx #1\expandafter \@firstoftwo
 \else \expandafter \@secondoftwo
 \fi
}%
\providecommand \natexlab [1]{#1}%
\providecommand \enquote  [1]{``#1''}%
\providecommand \bibnamefont  [1]{#1}%
\providecommand \bibfnamefont [1]{#1}%
\providecommand \citenamefont [1]{#1}%
\providecommand \href@noop [0]{\@secondoftwo}%
\providecommand \href [0]{\begingroup \@sanitize@url \@href}%
\providecommand \@href[1]{\@@startlink{#1}\@@href}%
\providecommand \@@href[1]{\endgroup#1\@@endlink}%
\providecommand \@sanitize@url [0]{\catcode `\\12\catcode `\$12\catcode
  `\&12\catcode `\#12\catcode `\^12\catcode `\_12\catcode `\%12\relax}%
\providecommand \@@startlink[1]{}%
\providecommand \@@endlink[0]{}%
\providecommand \url  [0]{\begingroup\@sanitize@url \@url }%
\providecommand \@url [1]{\endgroup\@href {#1}{\urlprefix }}%
\providecommand \urlprefix  [0]{URL }%
\providecommand \Eprint [0]{\href }%
\providecommand \doibase [0]{https://doi.org/}%
\providecommand \selectlanguage [0]{\@gobble}%
\providecommand \bibinfo  [0]{\@secondoftwo}%
\providecommand \bibfield  [0]{\@secondoftwo}%
\providecommand \translation [1]{[#1]}%
\providecommand \BibitemOpen [0]{}%
\providecommand \bibitemStop [0]{}%
\providecommand \bibitemNoStop [0]{.\EOS\space}%
\providecommand \EOS [0]{\spacefactor3000\relax}%
\providecommand \BibitemShut  [1]{\csname bibitem#1\endcsname}%
\let\auto@bib@innerbib\@empty
%</preamble>
\bibitem [{\citenamefont {Wigner}(1939)}]{wigner1939unitary}%
  \BibitemOpen
  \bibfield  {author} {\bibinfo {author} {\bibfnamefont {E.}~\bibnamefont
  {Wigner}},\ }\bibfield  {title} {\bibinfo {title} {On unitary representations
  of the inhomogeneous lorentz group},\ }\href
  {https://doi.org/10.2307/1968551} {\bibfield  {journal} {\bibinfo  {journal}
  {Annals of mathematics}\ ,\ \bibinfo {pages} {149}} (\bibinfo {year}
  {1939})}\BibitemShut {NoStop}%
\bibitem [{\citenamefont {Bargmann}(1947)}]{bargmann1947irreducible}%
  \BibitemOpen
  \bibfield  {author} {\bibinfo {author} {\bibfnamefont {V.}~\bibnamefont
  {Bargmann}},\ }\bibfield  {title} {\bibinfo {title} {Irreducible unitary
  representations of the lorentz group},\ }\href
  {https://doi.org/10.2307/1969129} {\bibfield  {journal} {\bibinfo  {journal}
  {Annals of Mathematics}\ ,\ \bibinfo {pages} {568}} (\bibinfo {year}
  {1947})}\BibitemShut {NoStop}%
\bibitem [{\citenamefont {Bargmann}\ and\ \citenamefont
  {Wigner}(1948)}]{bargmann1948group}%
  \BibitemOpen
  \bibfield  {author} {\bibinfo {author} {\bibfnamefont {V.}~\bibnamefont
  {Bargmann}}\ and\ \bibinfo {author} {\bibfnamefont {E.~P.}\ \bibnamefont
  {Wigner}},\ }\bibfield  {title} {\bibinfo {title} {Group theoretical
  discussion of relativistic wave equations},\ }\href
  {https://doi.org/10.1073/pnas.34.5.211} {\bibfield  {journal} {\bibinfo
  {journal} {Proceedings of the National Academy of Sciences}\ }\textbf
  {\bibinfo {volume} {34}},\ \bibinfo {pages} {211} (\bibinfo {year}
  {1948})}\BibitemShut {NoStop}%
\bibitem [{\citenamefont {Shirokov}(1958{\natexlab{a}})}]{shirokov1958group1}%
  \BibitemOpen
  \bibfield  {author} {\bibinfo {author} {\bibfnamefont {I.~M.}\ \bibnamefont
  {Shirokov}},\ }\bibfield  {title} {\bibinfo {title} {A group-theoretical
  consideration on the basis of relativistic quantum mechanics. i. the general
  properties of the inhomogeneous lorentz group},\ }\href
  {http://jetp.ras.ru/cgi-bin/e/index/e/6/4/p664?a=list} {\bibfield  {journal}
  {\bibinfo  {journal} {Journal of Experimental and Theoretical Physics}\
  }\textbf {\bibinfo {volume} {6}},\ \bibinfo {pages} {664} (\bibinfo {year}
  {1958}{\natexlab{a}})}\BibitemShut {NoStop}%
\bibitem [{\citenamefont {Shirokov}(1958{\natexlab{b}})}]{shirokov1958group2}%
  \BibitemOpen
  \bibfield  {author} {\bibinfo {author} {\bibfnamefont {I.~M.}\ \bibnamefont
  {Shirokov}},\ }\bibfield  {title} {\bibinfo {title} {A group-theoretical
  consideration on the basis of relativistic quantum mechanics. ii.
  classification of the irreducible representations of the inhomogeneous
  lorentz group},\ }\href
  {http://jetp.ras.ru/cgi-bin/e/index/e/6/5/p919?a=list} {\bibfield  {journal}
  {\bibinfo  {journal} {Journal of Experimental and Theoretical Physics}\
  }\textbf {\bibinfo {volume} {6}},\ \bibinfo {pages} {919} (\bibinfo {year}
  {1958}{\natexlab{b}})}\BibitemShut {NoStop}%
\bibitem [{\citenamefont {Shirokov}(1958{\natexlab{c}})}]{shirokov1958group3}%
  \BibitemOpen
  \bibfield  {author} {\bibinfo {author} {\bibfnamefont {I.~M.}\ \bibnamefont
  {Shirokov}},\ }\bibfield  {title} {\bibinfo {title} {A group-theoretical
  consideration on the basis of relativistic quantum mechanics. iii.
  irreducible representations of the classes p0 and o0, and the
  non-completely-reducible representations of the inhomogeneous lorentz
  group},\ }\href {http://jetp.ras.ru/cgi-bin/e/index/e/6/5/p929?a=list}
  {\bibfield  {journal} {\bibinfo  {journal} {Journal of Experimental and
  Theoretical Physics}\ }\textbf {\bibinfo {volume} {6}},\ \bibinfo {pages}
  {929} (\bibinfo {year} {1958}{\natexlab{c}})}\BibitemShut {NoStop}%
\bibitem [{\citenamefont {Shirokov}(1959)}]{shirokov1959relativistic}%
  \BibitemOpen
  \bibfield  {author} {\bibinfo {author} {\bibfnamefont {I.~M.}\ \bibnamefont
  {Shirokov}},\ }\bibfield  {title} {\bibinfo {title} {Relativistic theory of
  polarization effects},\ }\href
  {http://jetp.ras.ru/cgi-bin/e/index/e/8/4/p703?a=list} {\bibfield  {journal}
  {\bibinfo  {journal} {Journal of Experimental and Theoretical Physics}\
  }\textbf {\bibinfo {volume} {8}},\ \bibinfo {pages} {703} (\bibinfo {year}
  {1959})}\BibitemShut {NoStop}%
\bibitem [{\citenamefont {Tolhoek}(1956)}]{tolhoek1956electron}%
  \BibitemOpen
  \bibfield  {author} {\bibinfo {author} {\bibfnamefont {H.}~\bibnamefont
  {Tolhoek}},\ }\bibfield  {title} {\bibinfo {title} {Electron polarization,
  theory and experiment},\ }\href@noop {} {\bibfield  {journal} {\bibinfo
  {journal} {Reviews of modern physics}\ }\textbf {\bibinfo {volume} {28}},\
  \bibinfo {pages} {277} (\bibinfo {year} {1956})}\BibitemShut {NoStop}%
\bibitem [{\citenamefont {Anthony}\ \emph {et~al.}(1993)\citenamefont
  {Anthony}, \citenamefont {Arnold}, \citenamefont {Band}, \citenamefont
  {Borel}, \citenamefont {Bosted}, \citenamefont {Breton}, \citenamefont
  {Cates}, \citenamefont {Chupp}, \citenamefont {Dietrich}, \citenamefont
  {Dunne} \emph {et~al.}}]{anthony1993determination}%
  \BibitemOpen
  \bibfield  {author} {\bibinfo {author} {\bibfnamefont {P.}~\bibnamefont
  {Anthony}}, \bibinfo {author} {\bibfnamefont {R.}~\bibnamefont {Arnold}},
  \bibinfo {author} {\bibfnamefont {H.}~\bibnamefont {Band}}, \bibinfo {author}
  {\bibfnamefont {H.}~\bibnamefont {Borel}}, \bibinfo {author} {\bibfnamefont
  {P.}~\bibnamefont {Bosted}}, \bibinfo {author} {\bibfnamefont
  {V.}~\bibnamefont {Breton}}, \bibinfo {author} {\bibfnamefont
  {G.}~\bibnamefont {Cates}}, \bibinfo {author} {\bibfnamefont
  {T.}~\bibnamefont {Chupp}}, \bibinfo {author} {\bibfnamefont
  {F.}~\bibnamefont {Dietrich}}, \bibinfo {author} {\bibfnamefont
  {J.}~\bibnamefont {Dunne}}, \emph {et~al.},\ }\bibfield  {title} {\bibinfo
  {title} {Determination of the neutron spin structure function},\ }\href@noop
  {} {\bibfield  {journal} {\bibinfo  {journal} {Physical Review Letters}\
  }\textbf {\bibinfo {volume} {71}},\ \bibinfo {pages} {959} (\bibinfo {year}
  {1993})}\BibitemShut {NoStop}%
\bibitem [{\citenamefont {Abe}\ \emph {et~al.}(1995)\citenamefont {Abe},
  \citenamefont {Akagi}, \citenamefont {Anthony}, \citenamefont {Antonov},
  \citenamefont {Arnold}, \citenamefont {Averett}, \citenamefont {Band},
  \citenamefont {Bauer}, \citenamefont {Borel}, \citenamefont {Bosted} \emph
  {et~al.}}]{abe1995precision}%
  \BibitemOpen
  \bibfield  {author} {\bibinfo {author} {\bibfnamefont {K.}~\bibnamefont
  {Abe}}, \bibinfo {author} {\bibfnamefont {T.}~\bibnamefont {Akagi}}, \bibinfo
  {author} {\bibfnamefont {P.}~\bibnamefont {Anthony}}, \bibinfo {author}
  {\bibfnamefont {R.}~\bibnamefont {Antonov}}, \bibinfo {author} {\bibfnamefont
  {R.}~\bibnamefont {Arnold}}, \bibinfo {author} {\bibfnamefont
  {T.}~\bibnamefont {Averett}}, \bibinfo {author} {\bibfnamefont
  {H.}~\bibnamefont {Band}}, \bibinfo {author} {\bibfnamefont {J.}~\bibnamefont
  {Bauer}}, \bibinfo {author} {\bibfnamefont {H.}~\bibnamefont {Borel}},
  \bibinfo {author} {\bibfnamefont {P.}~\bibnamefont {Bosted}}, \emph
  {et~al.},\ }\bibfield  {title} {\bibinfo {title} {Precision measurement of
  the deuteron spin structure function g 1 d},\ }\href@noop {} {\bibfield
  {journal} {\bibinfo  {journal} {Physical Review Letters}\ }\textbf {\bibinfo
  {volume} {75}},\ \bibinfo {pages} {25} (\bibinfo {year} {1995})}\BibitemShut
  {NoStop}%
\bibitem [{\citenamefont {Prescott}\ \emph {et~al.}(1978)\citenamefont
  {Prescott}, \citenamefont {Atwood}, \citenamefont {Cottrell}, \citenamefont
  {Destaebler}, \citenamefont {Garwin}, \citenamefont {Gonidec}, \citenamefont
  {Miller}, \citenamefont {Rochester}, \citenamefont {Sato}, \citenamefont
  {Sherden} \emph {et~al.}}]{prescott1978parity}%
  \BibitemOpen
  \bibfield  {author} {\bibinfo {author} {\bibfnamefont {C.~Y.}\ \bibnamefont
  {Prescott}}, \bibinfo {author} {\bibfnamefont {W.}~\bibnamefont {Atwood}},
  \bibinfo {author} {\bibfnamefont {R.}~\bibnamefont {Cottrell}}, \bibinfo
  {author} {\bibfnamefont {H.}~\bibnamefont {Destaebler}}, \bibinfo {author}
  {\bibfnamefont {E.~L.}\ \bibnamefont {Garwin}}, \bibinfo {author}
  {\bibfnamefont {A.}~\bibnamefont {Gonidec}}, \bibinfo {author} {\bibfnamefont
  {R.~H.}\ \bibnamefont {Miller}}, \bibinfo {author} {\bibfnamefont
  {L.}~\bibnamefont {Rochester}}, \bibinfo {author} {\bibfnamefont
  {T.}~\bibnamefont {Sato}}, \bibinfo {author} {\bibfnamefont {D.}~\bibnamefont
  {Sherden}}, \emph {et~al.},\ }\bibfield  {title} {\bibinfo {title} {Parity
  non-conservation in inelastic electron scattering},\ }\href@noop {}
  {\bibfield  {journal} {\bibinfo  {journal} {Physics Letters B}\ }\textbf
  {\bibinfo {volume} {77}},\ \bibinfo {pages} {347} (\bibinfo {year}
  {1978})}\BibitemShut {NoStop}%
\bibitem [{\citenamefont {Anthony}\ \emph {et~al.}(2004)\citenamefont
  {Anthony}, \citenamefont {Arnold}, \citenamefont {Arroyo}, \citenamefont
  {Baird}, \citenamefont {Bega}, \citenamefont {Biesiada}, \citenamefont
  {Bosted}, \citenamefont {Breuer}, \citenamefont {Carr}, \citenamefont {Cates}
  \emph {et~al.}}]{anthony2004observation}%
  \BibitemOpen
  \bibfield  {author} {\bibinfo {author} {\bibfnamefont {P.}~\bibnamefont
  {Anthony}}, \bibinfo {author} {\bibfnamefont {R.}~\bibnamefont {Arnold}},
  \bibinfo {author} {\bibfnamefont {C.}~\bibnamefont {Arroyo}}, \bibinfo
  {author} {\bibfnamefont {K.}~\bibnamefont {Baird}}, \bibinfo {author}
  {\bibfnamefont {K.}~\bibnamefont {Bega}}, \bibinfo {author} {\bibfnamefont
  {J.}~\bibnamefont {Biesiada}}, \bibinfo {author} {\bibfnamefont
  {P.}~\bibnamefont {Bosted}}, \bibinfo {author} {\bibfnamefont
  {M.}~\bibnamefont {Breuer}}, \bibinfo {author} {\bibfnamefont
  {R.}~\bibnamefont {Carr}}, \bibinfo {author} {\bibfnamefont {G.}~\bibnamefont
  {Cates}}, \emph {et~al.},\ }\bibfield  {title} {\bibinfo {title} {Observation
  of parity nonconservation in m{\o}ller scattering},\ }\href@noop {}
  {\bibfield  {journal} {\bibinfo  {journal} {Physical review letters}\
  }\textbf {\bibinfo {volume} {92}},\ \bibinfo {pages} {181602} (\bibinfo
  {year} {2004})}\BibitemShut {NoStop}%
\bibitem [{\citenamefont {Collaboration}\ \emph {et~al.}(2018)\citenamefont
  {Collaboration} \emph {et~al.}}]{jefferson2018precision}%
  \BibitemOpen
  \bibfield  {author} {\bibinfo {author} {\bibfnamefont {J.~L.~Q.}\
  \bibnamefont {Collaboration}} \emph {et~al.},\ }\bibfield  {title} {\bibinfo
  {title} {Precision measurement of the weak charge of the proton},\
  }\href@noop {} {\bibfield  {journal} {\bibinfo  {journal} {Nature}\ }\textbf
  {\bibinfo {volume} {557}},\ \bibinfo {pages} {207} (\bibinfo {year}
  {2018})}\BibitemShut {NoStop}%
\bibitem [{\citenamefont {Moortgat-Pick}\ \emph {et~al.}(2008)\citenamefont
  {Moortgat-Pick}, \citenamefont {Abe}, \citenamefont {Alexander},
  \citenamefont {Ananthanarayan}, \citenamefont {Babich}, \citenamefont
  {Bharadwaj}, \citenamefont {Barber}, \citenamefont {Bartl}, \citenamefont
  {Brachmann}, \citenamefont {Chen} \emph {et~al.}}]{moortgat2008polarized}%
  \BibitemOpen
  \bibfield  {author} {\bibinfo {author} {\bibfnamefont {G.}~\bibnamefont
  {Moortgat-Pick}}, \bibinfo {author} {\bibfnamefont {T.}~\bibnamefont {Abe}},
  \bibinfo {author} {\bibfnamefont {G.}~\bibnamefont {Alexander}}, \bibinfo
  {author} {\bibfnamefont {B.}~\bibnamefont {Ananthanarayan}}, \bibinfo
  {author} {\bibfnamefont {A.}~\bibnamefont {Babich}}, \bibinfo {author}
  {\bibfnamefont {V.}~\bibnamefont {Bharadwaj}}, \bibinfo {author}
  {\bibfnamefont {D.}~\bibnamefont {Barber}}, \bibinfo {author} {\bibfnamefont
  {A.}~\bibnamefont {Bartl}}, \bibinfo {author} {\bibfnamefont
  {A.}~\bibnamefont {Brachmann}}, \bibinfo {author} {\bibfnamefont
  {S.}~\bibnamefont {Chen}}, \emph {et~al.},\ }\bibfield  {title} {\bibinfo
  {title} {Polarized positrons and electrons at the linear collider},\
  }\href@noop {} {\bibfield  {journal} {\bibinfo  {journal} {Physics Reports}\
  }\textbf {\bibinfo {volume} {460}},\ \bibinfo {pages} {131} (\bibinfo {year}
  {2008})}\BibitemShut {NoStop}%
\bibitem [{\citenamefont {Olsen}\ and\ \citenamefont
  {Maximon}(1959)}]{olsen1959photon}%
  \BibitemOpen
  \bibfield  {author} {\bibinfo {author} {\bibfnamefont {H.}~\bibnamefont
  {Olsen}}\ and\ \bibinfo {author} {\bibfnamefont {L.}~\bibnamefont
  {Maximon}},\ }\bibfield  {title} {\bibinfo {title} {Photon and electron
  polarization in high-energy bremsstrahlung and pair production with
  screening},\ }\href@noop {} {\bibfield  {journal} {\bibinfo  {journal}
  {Physical Review}\ }\textbf {\bibinfo {volume} {114}},\ \bibinfo {pages}
  {887} (\bibinfo {year} {1959})}\BibitemShut {NoStop}%
\bibitem [{\citenamefont {M{\"a}rtin}\ \emph {et~al.}(2012)\citenamefont
  {M{\"a}rtin}, \citenamefont {Weber}, \citenamefont {Barday}, \citenamefont
  {Fritzsche}, \citenamefont {Spillmann}, \citenamefont {Chen}, \citenamefont
  {DuBois}, \citenamefont {Enders}, \citenamefont {Hegewald}, \citenamefont
  {Hess} \emph {et~al.}}]{martin2012polarization}%
  \BibitemOpen
  \bibfield  {author} {\bibinfo {author} {\bibfnamefont {R.}~\bibnamefont
  {M{\"a}rtin}}, \bibinfo {author} {\bibfnamefont {G.}~\bibnamefont {Weber}},
  \bibinfo {author} {\bibfnamefont {R.}~\bibnamefont {Barday}}, \bibinfo
  {author} {\bibfnamefont {Y.}~\bibnamefont {Fritzsche}}, \bibinfo {author}
  {\bibfnamefont {U.}~\bibnamefont {Spillmann}}, \bibinfo {author}
  {\bibfnamefont {W.}~\bibnamefont {Chen}}, \bibinfo {author} {\bibfnamefont
  {R.}~\bibnamefont {DuBois}}, \bibinfo {author} {\bibfnamefont
  {J.}~\bibnamefont {Enders}}, \bibinfo {author} {\bibfnamefont
  {M.}~\bibnamefont {Hegewald}}, \bibinfo {author} {\bibfnamefont
  {S.}~\bibnamefont {Hess}}, \emph {et~al.},\ }\bibfield  {title} {\bibinfo
  {title} {Polarization transfer of bremsstrahlung arising from spin-polarized
  electrons},\ }\href@noop {} {\bibfield  {journal} {\bibinfo  {journal}
  {Physical review letters}\ }\textbf {\bibinfo {volume} {108}},\ \bibinfo
  {pages} {264801} (\bibinfo {year} {2012})}\BibitemShut {NoStop}%
\bibitem [{\citenamefont {Abbott}\ \emph {et~al.}(2016)\citenamefont {Abbott},
  \citenamefont {Adderley}, \citenamefont {Adeyemi}, \citenamefont {Aguilera},
  \citenamefont {Ali}, \citenamefont {Areti}, \citenamefont {Baylac},
  \citenamefont {Benesch}, \citenamefont {Bosson}, \citenamefont {Cade} \emph
  {et~al.}}]{abbott2016production}%
  \BibitemOpen
  \bibfield  {author} {\bibinfo {author} {\bibfnamefont {D.}~\bibnamefont
  {Abbott}}, \bibinfo {author} {\bibfnamefont {P.}~\bibnamefont {Adderley}},
  \bibinfo {author} {\bibfnamefont {A.}~\bibnamefont {Adeyemi}}, \bibinfo
  {author} {\bibfnamefont {P.}~\bibnamefont {Aguilera}}, \bibinfo {author}
  {\bibfnamefont {M.}~\bibnamefont {Ali}}, \bibinfo {author} {\bibfnamefont
  {H.}~\bibnamefont {Areti}}, \bibinfo {author} {\bibfnamefont
  {M.}~\bibnamefont {Baylac}}, \bibinfo {author} {\bibfnamefont
  {J.}~\bibnamefont {Benesch}}, \bibinfo {author} {\bibfnamefont
  {G.}~\bibnamefont {Bosson}}, \bibinfo {author} {\bibfnamefont
  {B.}~\bibnamefont {Cade}}, \emph {et~al.},\ }\bibfield  {title} {\bibinfo
  {title} {Production of highly polarized positrons using polarized electrons
  at mev energies},\ }\href@noop {} {\bibfield  {journal} {\bibinfo  {journal}
  {Physical review letters}\ }\textbf {\bibinfo {volume} {116}},\ \bibinfo
  {pages} {214801} (\bibinfo {year} {2016})}\BibitemShut {NoStop}%
\bibitem [{\citenamefont {Mane}\ \emph {et~al.}(2005)\citenamefont {Mane},
  \citenamefont {Shatunov},\ and\ \citenamefont {Yokoya}}]{mane2005spin}%
  \BibitemOpen
  \bibfield  {author} {\bibinfo {author} {\bibfnamefont {S.}~\bibnamefont
  {Mane}}, \bibinfo {author} {\bibfnamefont {Y.~M.}\ \bibnamefont {Shatunov}},\
  and\ \bibinfo {author} {\bibfnamefont {K.}~\bibnamefont {Yokoya}},\
  }\bibfield  {title} {\bibinfo {title} {Spin-polarized charged particle beams
  in high-energy accelerators},\ }\href@noop {} {\bibfield  {journal} {\bibinfo
   {journal} {Reports on Progress in Physics}\ }\textbf {\bibinfo {volume}
  {68}},\ \bibinfo {pages} {1997} (\bibinfo {year} {2005})}\BibitemShut
  {NoStop}%
\bibitem [{\citenamefont {Sokolov}\ and\ \citenamefont
  {Ternov}(1967)}]{sokolov1967synchrotron}%
  \BibitemOpen
  \bibfield  {author} {\bibinfo {author} {\bibfnamefont {A.}~\bibnamefont
  {Sokolov}}\ and\ \bibinfo {author} {\bibfnamefont {I.}~\bibnamefont
  {Ternov}},\ }\bibfield  {title} {\bibinfo {title} {Synchrotron radiation},\
  }\href@noop {} {\bibfield  {journal} {\bibinfo  {journal} {Soviet Physics
  Journal}\ }\textbf {\bibinfo {volume} {10}},\ \bibinfo {pages} {39} (\bibinfo
  {year} {1967})}\BibitemShut {NoStop}%
\bibitem [{\citenamefont {Bordovitsyn}(1999)}]{bordovitsyn1999synchrotron}%
  \BibitemOpen
  \bibfield  {author} {\bibinfo {author} {\bibfnamefont {V.~A.}\ \bibnamefont
  {Bordovitsyn}},\ }\href {https://doi.org/10.1142/3492} {\emph {\bibinfo
  {title} {Synchrotron radiation theory and its development: in memory of I M
  Ternov}}}\ (\bibinfo  {publisher} {World Scientific},\ \bibinfo {address}
  {Singapore},\ \bibinfo {year} {1999})\BibitemShut {NoStop}%
\bibitem [{\citenamefont {Sun}\ \emph {et~al.}(2010)\citenamefont {Sun},
  \citenamefont {Zhang}, \citenamefont {Li}, \citenamefont {Wu}, \citenamefont
  {Mikhailov}, \citenamefont {Popov}, \citenamefont {Xu}, \citenamefont
  {Chao},\ and\ \citenamefont {Wu}}]{sun2010polarization}%
  \BibitemOpen
  \bibfield  {author} {\bibinfo {author} {\bibfnamefont {C.}~\bibnamefont
  {Sun}}, \bibinfo {author} {\bibfnamefont {J.}~\bibnamefont {Zhang}}, \bibinfo
  {author} {\bibfnamefont {J.}~\bibnamefont {Li}}, \bibinfo {author}
  {\bibfnamefont {W.}~\bibnamefont {Wu}}, \bibinfo {author} {\bibfnamefont
  {S.}~\bibnamefont {Mikhailov}}, \bibinfo {author} {\bibfnamefont
  {V.}~\bibnamefont {Popov}}, \bibinfo {author} {\bibfnamefont
  {H.}~\bibnamefont {Xu}}, \bibinfo {author} {\bibfnamefont {A.}~\bibnamefont
  {Chao}},\ and\ \bibinfo {author} {\bibfnamefont {Y.}~\bibnamefont {Wu}},\
  }\bibfield  {title} {\bibinfo {title} {Polarization measurement of stored
  electron beam using touschek lifetime},\ }\href@noop {} {\bibfield  {journal}
  {\bibinfo  {journal} {Nuclear Instruments and Methods in Physics Research
  Section A: Accelerators, Spectrometers, Detectors and Associated Equipment}\
  }\textbf {\bibinfo {volume} {614}},\ \bibinfo {pages} {339} (\bibinfo {year}
  {2010})}\BibitemShut {NoStop}%
\bibitem [{\citenamefont {Buon}\ and\ \citenamefont
  {Steffen}(1986)}]{buon1986hera}%
  \BibitemOpen
  \bibfield  {author} {\bibinfo {author} {\bibfnamefont {J.}~\bibnamefont
  {Buon}}\ and\ \bibinfo {author} {\bibfnamefont {K.}~\bibnamefont {Steffen}},\
  }\bibfield  {title} {\bibinfo {title} {Hera variable-energy “mini” spin
  rotator and head-on ep collision scheme with choice of electron helicity},\
  }\href@noop {} {\bibfield  {journal} {\bibinfo  {journal} {Nuclear
  Instruments and Methods in Physics Research Section A: Accelerators,
  Spectrometers, Detectors and Associated Equipment}\ }\textbf {\bibinfo
  {volume} {245}},\ \bibinfo {pages} {248} (\bibinfo {year}
  {1986})}\BibitemShut {NoStop}%
\bibitem [{\citenamefont {Pierce}\ \emph {et~al.}(1975)\citenamefont {Pierce},
  \citenamefont {Meier},\ and\ \citenamefont
  {Z{\"u}rcher}}]{pierce1975negative}%
  \BibitemOpen
  \bibfield  {author} {\bibinfo {author} {\bibfnamefont {D.~T.}\ \bibnamefont
  {Pierce}}, \bibinfo {author} {\bibfnamefont {F.}~\bibnamefont {Meier}},\ and\
  \bibinfo {author} {\bibfnamefont {P.}~\bibnamefont {Z{\"u}rcher}},\
  }\bibfield  {title} {\bibinfo {title} {Negative electron affinity gaas: A new
  source of spin-polarized electrons},\ }\href@noop {} {\bibfield  {journal}
  {\bibinfo  {journal} {Applied Physics Letters}\ }\textbf {\bibinfo {volume}
  {26}},\ \bibinfo {pages} {670} (\bibinfo {year} {1975})}\BibitemShut
  {NoStop}%
\bibitem [{\citenamefont {Pierce}\ and\ \citenamefont
  {Meier}(1976)}]{pierce1976photoemission}%
  \BibitemOpen
  \bibfield  {author} {\bibinfo {author} {\bibfnamefont {D.~T.}\ \bibnamefont
  {Pierce}}\ and\ \bibinfo {author} {\bibfnamefont {F.}~\bibnamefont {Meier}},\
  }\bibfield  {title} {\bibinfo {title} {Photoemission of spin-polarized
  electrons from gaas},\ }\href@noop {} {\bibfield  {journal} {\bibinfo
  {journal} {Physical Review B}\ }\textbf {\bibinfo {volume} {13}},\ \bibinfo
  {pages} {5484} (\bibinfo {year} {1976})}\BibitemShut {NoStop}%
\bibitem [{\citenamefont {Batelaan}\ \emph {et~al.}(1999)\citenamefont
  {Batelaan}, \citenamefont {Green}, \citenamefont {Hitt},\ and\ \citenamefont
  {Gay}}]{batelaan1999optically}%
  \BibitemOpen
  \bibfield  {author} {\bibinfo {author} {\bibfnamefont {H.}~\bibnamefont
  {Batelaan}}, \bibinfo {author} {\bibfnamefont {A.}~\bibnamefont {Green}},
  \bibinfo {author} {\bibfnamefont {B.}~\bibnamefont {Hitt}},\ and\ \bibinfo
  {author} {\bibfnamefont {T.~J.}\ \bibnamefont {Gay}},\ }\bibfield  {title}
  {\bibinfo {title} {Optically pumped electron spin filter},\ }\href@noop {}
  {\bibfield  {journal} {\bibinfo  {journal} {Physical review letters}\
  }\textbf {\bibinfo {volume} {82}},\ \bibinfo {pages} {4216} (\bibinfo {year}
  {1999})}\BibitemShut {NoStop}%
\bibitem [{\citenamefont {Jackson}(1998)}]{jackson1998classical}%
  \BibitemOpen
  \bibfield  {author} {\bibinfo {author} {\bibfnamefont {J.~D.}\ \bibnamefont
  {Jackson}},\ }\href@noop {} {\emph {\bibinfo {title} {Classical
  electrodynamics}}},\ \bibinfo {edition} {3rd}\ ed.\ (\bibinfo  {publisher}
  {John Wiley \& Sons},\ \bibinfo {address} {Hoboken},\ \bibinfo {year}
  {1998})\BibitemShut {NoStop}%
\bibitem [{\citenamefont {Steiner}\ \emph {et~al.}(2007)\citenamefont
  {Steiner}, \citenamefont {Ackermann}, \citenamefont {Muller},\ and\
  \citenamefont {Weiland}}]{steiner2007wien}%
  \BibitemOpen
  \bibfield  {author} {\bibinfo {author} {\bibfnamefont {B.}~\bibnamefont
  {Steiner}}, \bibinfo {author} {\bibfnamefont {W.}~\bibnamefont {Ackermann}},
  \bibinfo {author} {\bibfnamefont {W.}~\bibnamefont {Muller}},\ and\ \bibinfo
  {author} {\bibfnamefont {T.}~\bibnamefont {Weiland}},\ }\bibfield  {title}
  {\bibinfo {title} {Wien filter as a spin rotator at low energy},\ }in\
  \href@noop {} {\emph {\bibinfo {booktitle} {2007 IEEE Particle Accelerator
  Conference (PAC)}}}\ (\bibinfo {organization} {IEEE},\ \bibinfo {year}
  {2007})\ pp.\ \bibinfo {pages} {170--172}\BibitemShut {NoStop}%
\bibitem [{\citenamefont {Mott}(1929)}]{mott1929scattering}%
  \BibitemOpen
  \bibfield  {author} {\bibinfo {author} {\bibfnamefont {N.~F.}\ \bibnamefont
  {Mott}},\ }\bibfield  {title} {\bibinfo {title} {The scattering of fast
  electrons by atomic nuclei},\ }\href@noop {} {\bibfield  {journal} {\bibinfo
  {journal} {Proceedings of the Royal Society of London. Series A, Containing
  Papers of a Mathematical and Physical Character}\ }\textbf {\bibinfo {volume}
  {124}},\ \bibinfo {pages} {425} (\bibinfo {year} {1929})}\BibitemShut
  {NoStop}%
\bibitem [{\citenamefont {Gay}\ and\ \citenamefont
  {Dunning}(1992)}]{gay1992mott}%
  \BibitemOpen
  \bibfield  {author} {\bibinfo {author} {\bibfnamefont {T.~J.}\ \bibnamefont
  {Gay}}\ and\ \bibinfo {author} {\bibfnamefont {F.}~\bibnamefont {Dunning}},\
  }\bibfield  {title} {\bibinfo {title} {Mott electron polarimetry},\
  }\href@noop {} {\bibfield  {journal} {\bibinfo  {journal} {Review of
  scientific instruments}\ }\textbf {\bibinfo {volume} {63}},\ \bibinfo {pages}
  {1635} (\bibinfo {year} {1992})}\BibitemShut {NoStop}%
\bibitem [{\citenamefont {M{\o}ller}(1932)}]{moller1932theorie}%
  \BibitemOpen
  \bibfield  {author} {\bibinfo {author} {\bibfnamefont {C.}~\bibnamefont
  {M{\o}ller}},\ }\bibfield  {title} {\bibinfo {title} {Zur theorie des
  durchgangs schneller elektronen durch materie},\ }\href@noop {} {\bibfield
  {journal} {\bibinfo  {journal} {Annalen der Physik}\ }\textbf {\bibinfo
  {volume} {406}},\ \bibinfo {pages} {531} (\bibinfo {year}
  {1932})}\BibitemShut {NoStop}%
\bibitem [{\citenamefont {Cooper}\ \emph {et~al.}(1975)\citenamefont {Cooper},
  \citenamefont {Alguard}, \citenamefont {Ehrlich}, \citenamefont {Hughes},
  \citenamefont {Kobayakawa}, \citenamefont {Ladish}, \citenamefont {Lubell},
  \citenamefont {Sasao}, \citenamefont {Sch{\"u}ler}, \citenamefont {Souder}
  \emph {et~al.}}]{cooper1975polarized}%
  \BibitemOpen
  \bibfield  {author} {\bibinfo {author} {\bibfnamefont {P.}~\bibnamefont
  {Cooper}}, \bibinfo {author} {\bibfnamefont {M.}~\bibnamefont {Alguard}},
  \bibinfo {author} {\bibfnamefont {R.}~\bibnamefont {Ehrlich}}, \bibinfo
  {author} {\bibfnamefont {V.}~\bibnamefont {Hughes}}, \bibinfo {author}
  {\bibfnamefont {H.}~\bibnamefont {Kobayakawa}}, \bibinfo {author}
  {\bibfnamefont {J.}~\bibnamefont {Ladish}}, \bibinfo {author} {\bibfnamefont
  {M.}~\bibnamefont {Lubell}}, \bibinfo {author} {\bibfnamefont
  {N.}~\bibnamefont {Sasao}}, \bibinfo {author} {\bibfnamefont
  {K.}~\bibnamefont {Sch{\"u}ler}}, \bibinfo {author} {\bibfnamefont
  {P.}~\bibnamefont {Souder}}, \emph {et~al.},\ }\bibfield  {title} {\bibinfo
  {title} {Polarized electron-electron scattering at gev energies},\
  }\href@noop {} {\bibfield  {journal} {\bibinfo  {journal} {Physical Review
  Letters}\ }\textbf {\bibinfo {volume} {34}},\ \bibinfo {pages} {1589}
  (\bibinfo {year} {1975})}\BibitemShut {NoStop}%
\bibitem [{\citenamefont {Hauger}\ \emph {et~al.}(2001)\citenamefont {Hauger},
  \citenamefont {Honegger}, \citenamefont {Jourdan}, \citenamefont {Kubon},
  \citenamefont {Petitjean}, \citenamefont {Rohe}, \citenamefont {Sick},
  \citenamefont {Warren}, \citenamefont {W{\"o}hrle}, \citenamefont {Zhao}
  \emph {et~al.}}]{hauger2001high}%
  \BibitemOpen
  \bibfield  {author} {\bibinfo {author} {\bibfnamefont {M.}~\bibnamefont
  {Hauger}}, \bibinfo {author} {\bibfnamefont {A.}~\bibnamefont {Honegger}},
  \bibinfo {author} {\bibfnamefont {J.}~\bibnamefont {Jourdan}}, \bibinfo
  {author} {\bibfnamefont {G.}~\bibnamefont {Kubon}}, \bibinfo {author}
  {\bibfnamefont {T.}~\bibnamefont {Petitjean}}, \bibinfo {author}
  {\bibfnamefont {D.}~\bibnamefont {Rohe}}, \bibinfo {author} {\bibfnamefont
  {I.}~\bibnamefont {Sick}}, \bibinfo {author} {\bibfnamefont {G.}~\bibnamefont
  {Warren}}, \bibinfo {author} {\bibfnamefont {H.}~\bibnamefont {W{\"o}hrle}},
  \bibinfo {author} {\bibfnamefont {J.}~\bibnamefont {Zhao}}, \emph {et~al.},\
  }\bibfield  {title} {\bibinfo {title} {A high-precision polarimeter},\
  }\href@noop {} {\bibfield  {journal} {\bibinfo  {journal} {Nuclear
  Instruments and Methods in Physics Research Section A: Accelerators,
  Spectrometers, Detectors and Associated Equipment}\ }\textbf {\bibinfo
  {volume} {462}},\ \bibinfo {pages} {382} (\bibinfo {year}
  {2001})}\BibitemShut {NoStop}%
\bibitem [{\citenamefont {Barber}\ \emph {et~al.}(1993)\citenamefont {Barber},
  \citenamefont {Bremer}, \citenamefont {B{\"o}ge}, \citenamefont {Brinkmann},
  \citenamefont {Br{\"u}ckner}, \citenamefont {B{\"u}scher}, \citenamefont
  {Chapman}, \citenamefont {Coulter}, \citenamefont {Delheij}, \citenamefont
  {D{\"u}ren} \emph {et~al.}}]{barber1993hera}%
  \BibitemOpen
  \bibfield  {author} {\bibinfo {author} {\bibfnamefont {D.}~\bibnamefont
  {Barber}}, \bibinfo {author} {\bibfnamefont {H.-D.}\ \bibnamefont {Bremer}},
  \bibinfo {author} {\bibfnamefont {M.}~\bibnamefont {B{\"o}ge}}, \bibinfo
  {author} {\bibfnamefont {R.}~\bibnamefont {Brinkmann}}, \bibinfo {author}
  {\bibfnamefont {W.}~\bibnamefont {Br{\"u}ckner}}, \bibinfo {author}
  {\bibfnamefont {C.}~\bibnamefont {B{\"u}scher}}, \bibinfo {author}
  {\bibfnamefont {M.}~\bibnamefont {Chapman}}, \bibinfo {author} {\bibfnamefont
  {K.}~\bibnamefont {Coulter}}, \bibinfo {author} {\bibfnamefont
  {P.}~\bibnamefont {Delheij}}, \bibinfo {author} {\bibfnamefont
  {M.}~\bibnamefont {D{\"u}ren}}, \emph {et~al.},\ }\bibfield  {title}
  {\bibinfo {title} {The hera polarimeter and the first observation of electron
  spin polarization at hera},\ }\href@noop {} {\bibfield  {journal} {\bibinfo
  {journal} {Nuclear Instruments and Methods in Physics Research Section A:
  Accelerators, Spectrometers, Detectors and Associated Equipment}\ }\textbf
  {\bibinfo {volume} {329}},\ \bibinfo {pages} {79} (\bibinfo {year}
  {1993})}\BibitemShut {NoStop}%
\bibitem [{\citenamefont {Beckmann}\ \emph {et~al.}(2002)\citenamefont
  {Beckmann}, \citenamefont {Borissov}, \citenamefont {Brauksiepe},
  \citenamefont {Burkart}, \citenamefont {Fischer}, \citenamefont {Franz},
  \citenamefont {Heinsius}, \citenamefont {K{\"o}nigsmann}, \citenamefont
  {Lorenzon}, \citenamefont {Menden} \emph
  {et~al.}}]{beckmann2002longitudinal}%
  \BibitemOpen
  \bibfield  {author} {\bibinfo {author} {\bibfnamefont {M.}~\bibnamefont
  {Beckmann}}, \bibinfo {author} {\bibfnamefont {A.}~\bibnamefont {Borissov}},
  \bibinfo {author} {\bibfnamefont {S.}~\bibnamefont {Brauksiepe}}, \bibinfo
  {author} {\bibfnamefont {F.}~\bibnamefont {Burkart}}, \bibinfo {author}
  {\bibfnamefont {H.}~\bibnamefont {Fischer}}, \bibinfo {author} {\bibfnamefont
  {J.}~\bibnamefont {Franz}}, \bibinfo {author} {\bibfnamefont
  {F.}~\bibnamefont {Heinsius}}, \bibinfo {author} {\bibfnamefont
  {K.}~\bibnamefont {K{\"o}nigsmann}}, \bibinfo {author} {\bibfnamefont
  {W.}~\bibnamefont {Lorenzon}}, \bibinfo {author} {\bibfnamefont
  {F.}~\bibnamefont {Menden}}, \emph {et~al.},\ }\bibfield  {title} {\bibinfo
  {title} {The longitudinal polarimeter at hera},\ }\href@noop {} {\bibfield
  {journal} {\bibinfo  {journal} {Nuclear Instruments and Methods in Physics
  Research Section A: Accelerators, Spectrometers, Detectors and Associated
  Equipment}\ }\textbf {\bibinfo {volume} {479}},\ \bibinfo {pages} {334}
  (\bibinfo {year} {2002})}\BibitemShut {NoStop}%
\bibitem [{\citenamefont {Belomesthnykh}\ \emph {et~al.}(1984)\citenamefont
  {Belomesthnykh}, \citenamefont {Bondar}, \citenamefont {Yegorychev},
  \citenamefont {Zhilitch}, \citenamefont {Kornyukhin}, \citenamefont
  {Nikitin}, \citenamefont {Saldin}, \citenamefont {Skrinsky},\ and\
  \citenamefont {Tumaikin}}]{belomesthnykh1984observation}%
  \BibitemOpen
  \bibfield  {author} {\bibinfo {author} {\bibfnamefont {S.}~\bibnamefont
  {Belomesthnykh}}, \bibinfo {author} {\bibfnamefont {A.}~\bibnamefont
  {Bondar}}, \bibinfo {author} {\bibfnamefont {M.}~\bibnamefont {Yegorychev}},
  \bibinfo {author} {\bibfnamefont {V.}~\bibnamefont {Zhilitch}}, \bibinfo
  {author} {\bibfnamefont {G.}~\bibnamefont {Kornyukhin}}, \bibinfo {author}
  {\bibfnamefont {S.}~\bibnamefont {Nikitin}}, \bibinfo {author} {\bibfnamefont
  {E.}~\bibnamefont {Saldin}}, \bibinfo {author} {\bibfnamefont
  {A.}~\bibnamefont {Skrinsky}},\ and\ \bibinfo {author} {\bibfnamefont
  {G.}~\bibnamefont {Tumaikin}},\ }\bibfield  {title} {\bibinfo {title} {An
  observation of the spin dependence of synchrotron radiation intensity},\
  }\href@noop {} {\bibfield  {journal} {\bibinfo  {journal} {Nuclear
  Instruments and Methods in Physics Research Section A: Accelerators,
  Spectrometers, Detectors and Associated Equipment}\ }\textbf {\bibinfo
  {volume} {227}},\ \bibinfo {pages} {173} (\bibinfo {year}
  {1984})}\BibitemShut {NoStop}%
\bibitem [{\citenamefont {Danson}\ \emph {et~al.}(2019)\citenamefont {Danson},
  \citenamefont {Haefner}, \citenamefont {Bromage}, \citenamefont {Butcher},
  \citenamefont {Chanteloup}, \citenamefont {Chowdhury}, \citenamefont
  {Galvanauskas}, \citenamefont {Gizzi}, \citenamefont {Hein}, \citenamefont
  {Hillier} \emph {et~al.}}]{danson2019petawatt}%
  \BibitemOpen
  \bibfield  {author} {\bibinfo {author} {\bibfnamefont {C.~N.}\ \bibnamefont
  {Danson}}, \bibinfo {author} {\bibfnamefont {C.}~\bibnamefont {Haefner}},
  \bibinfo {author} {\bibfnamefont {J.}~\bibnamefont {Bromage}}, \bibinfo
  {author} {\bibfnamefont {T.}~\bibnamefont {Butcher}}, \bibinfo {author}
  {\bibfnamefont {J.-C.~F.}\ \bibnamefont {Chanteloup}}, \bibinfo {author}
  {\bibfnamefont {E.~A.}\ \bibnamefont {Chowdhury}}, \bibinfo {author}
  {\bibfnamefont {A.}~\bibnamefont {Galvanauskas}}, \bibinfo {author}
  {\bibfnamefont {L.~A.}\ \bibnamefont {Gizzi}}, \bibinfo {author}
  {\bibfnamefont {J.}~\bibnamefont {Hein}}, \bibinfo {author} {\bibfnamefont
  {D.~I.}\ \bibnamefont {Hillier}}, \emph {et~al.},\ }\bibfield  {title}
  {\bibinfo {title} {Petawatt and exawatt class lasers worldwide},\ }\href@noop
  {} {\bibfield  {journal} {\bibinfo  {journal} {High Power Laser Science and
  Engineering}\ }\textbf {\bibinfo {volume} {7}} (\bibinfo {year}
  {2019})}\BibitemShut {NoStop}%
\bibitem [{\citenamefont {Yoon}\ \emph {et~al.}(2019)\citenamefont {Yoon},
  \citenamefont {Jeon}, \citenamefont {Shin}, \citenamefont {Lee},
  \citenamefont {Lee}, \citenamefont {Choi}, \citenamefont {Kim}, \citenamefont
  {Sung},\ and\ \citenamefont {Nam}}]{yoon2019achieving}%
  \BibitemOpen
  \bibfield  {author} {\bibinfo {author} {\bibfnamefont {J.~W.}\ \bibnamefont
  {Yoon}}, \bibinfo {author} {\bibfnamefont {C.}~\bibnamefont {Jeon}}, \bibinfo
  {author} {\bibfnamefont {J.}~\bibnamefont {Shin}}, \bibinfo {author}
  {\bibfnamefont {S.~K.}\ \bibnamefont {Lee}}, \bibinfo {author} {\bibfnamefont
  {H.~W.}\ \bibnamefont {Lee}}, \bibinfo {author} {\bibfnamefont {I.~W.}\
  \bibnamefont {Choi}}, \bibinfo {author} {\bibfnamefont {H.~T.}\ \bibnamefont
  {Kim}}, \bibinfo {author} {\bibfnamefont {J.~H.}\ \bibnamefont {Sung}},\ and\
  \bibinfo {author} {\bibfnamefont {C.~H.}\ \bibnamefont {Nam}},\ }\bibfield
  {title} {\bibinfo {title} {Achieving the laser intensity of 5.5$\times$ 10 22
  w/cm 2 with a wavefront-corrected multi-pw laser},\ }\href@noop {} {\bibfield
   {journal} {\bibinfo  {journal} {Optics express}\ }\textbf {\bibinfo {volume}
  {27}},\ \bibinfo {pages} {20412} (\bibinfo {year} {2019})}\BibitemShut
  {NoStop}%
\bibitem [{\citenamefont {Yoon}\ \emph {et~al.}(2021)\citenamefont {Yoon},
  \citenamefont {Kim}, \citenamefont {Choi}, \citenamefont {Sung},
  \citenamefont {Lee}, \citenamefont {Lee},\ and\ \citenamefont
  {Nam}}]{yoon2021realization}%
  \BibitemOpen
  \bibfield  {author} {\bibinfo {author} {\bibfnamefont {J.~W.}\ \bibnamefont
  {Yoon}}, \bibinfo {author} {\bibfnamefont {Y.~G.}\ \bibnamefont {Kim}},
  \bibinfo {author} {\bibfnamefont {I.~W.}\ \bibnamefont {Choi}}, \bibinfo
  {author} {\bibfnamefont {J.~H.}\ \bibnamefont {Sung}}, \bibinfo {author}
  {\bibfnamefont {H.~W.}\ \bibnamefont {Lee}}, \bibinfo {author} {\bibfnamefont
  {S.~K.}\ \bibnamefont {Lee}},\ and\ \bibinfo {author} {\bibfnamefont {C.~H.}\
  \bibnamefont {Nam}},\ }\bibfield  {title} {\bibinfo {title} {Realization of
  laser intensity over 10 23 w/cm 2},\ }\href@noop {} {\bibfield  {journal}
  {\bibinfo  {journal} {Optica}\ }\textbf {\bibinfo {volume} {8}},\ \bibinfo
  {pages} {630} (\bibinfo {year} {2021})}\BibitemShut {NoStop}%
\bibitem [{\citenamefont {Li}\ \emph {et~al.}(2019)\citenamefont {Li},
  \citenamefont {Shaisultanov}, \citenamefont {Hatsagortsyan}, \citenamefont
  {Wan}, \citenamefont {Keitel},\ and\ \citenamefont
  {Li}}]{li2019ultrarelativistic}%
  \BibitemOpen
  \bibfield  {author} {\bibinfo {author} {\bibfnamefont {Y.-F.}\ \bibnamefont
  {Li}}, \bibinfo {author} {\bibfnamefont {R.}~\bibnamefont {Shaisultanov}},
  \bibinfo {author} {\bibfnamefont {K.~Z.}\ \bibnamefont {Hatsagortsyan}},
  \bibinfo {author} {\bibfnamefont {F.}~\bibnamefont {Wan}}, \bibinfo {author}
  {\bibfnamefont {C.~H.}\ \bibnamefont {Keitel}},\ and\ \bibinfo {author}
  {\bibfnamefont {J.-X.}\ \bibnamefont {Li}},\ }\bibfield  {title} {\bibinfo
  {title} {Ultrarelativistic electron-beam polarization in single-shot
  interaction with an ultraintense laser pulse},\ }\href@noop {} {\bibfield
  {journal} {\bibinfo  {journal} {Physical review letters}\ }\textbf {\bibinfo
  {volume} {122}},\ \bibinfo {pages} {154801} (\bibinfo {year}
  {2019})}\BibitemShut {NoStop}%
\bibitem [{\citenamefont {Song}\ \emph {et~al.}(2019)\citenamefont {Song},
  \citenamefont {Wang}, \citenamefont {Li}, \citenamefont {Li},\ and\
  \citenamefont {Li}}]{song2019spin}%
  \BibitemOpen
  \bibfield  {author} {\bibinfo {author} {\bibfnamefont {H.-H.}\ \bibnamefont
  {Song}}, \bibinfo {author} {\bibfnamefont {W.-M.}\ \bibnamefont {Wang}},
  \bibinfo {author} {\bibfnamefont {J.-X.}\ \bibnamefont {Li}}, \bibinfo
  {author} {\bibfnamefont {Y.-F.}\ \bibnamefont {Li}},\ and\ \bibinfo {author}
  {\bibfnamefont {Y.-T.}\ \bibnamefont {Li}},\ }\bibfield  {title} {\bibinfo
  {title} {Spin-polarization effects of an ultrarelativistic electron beam in
  an ultraintense two-color laser pulse},\ }\href@noop {} {\bibfield  {journal}
  {\bibinfo  {journal} {Physical Review A}\ }\textbf {\bibinfo {volume}
  {100}},\ \bibinfo {pages} {033407} (\bibinfo {year} {2019})}\BibitemShut
  {NoStop}%
\bibitem [{\citenamefont {Seipt}\ \emph {et~al.}(2018)\citenamefont {Seipt},
  \citenamefont {Del~Sorbo}, \citenamefont {Ridgers},\ and\ \citenamefont
  {Thomas}}]{seipt2018theory}%
  \BibitemOpen
  \bibfield  {author} {\bibinfo {author} {\bibfnamefont {D.}~\bibnamefont
  {Seipt}}, \bibinfo {author} {\bibfnamefont {D.}~\bibnamefont {Del~Sorbo}},
  \bibinfo {author} {\bibfnamefont {C.}~\bibnamefont {Ridgers}},\ and\ \bibinfo
  {author} {\bibfnamefont {A.}~\bibnamefont {Thomas}},\ }\bibfield  {title}
  {\bibinfo {title} {Theory of radiative electron polarization in strong laser
  fields},\ }\href@noop {} {\bibfield  {journal} {\bibinfo  {journal} {Physical
  Review A}\ }\textbf {\bibinfo {volume} {98}},\ \bibinfo {pages} {023417}
  (\bibinfo {year} {2018})}\BibitemShut {NoStop}%
\bibitem [{\citenamefont {Seipt}\ \emph {et~al.}(2019)\citenamefont {Seipt},
  \citenamefont {Del~Sorbo}, \citenamefont {Ridgers},\ and\ \citenamefont
  {Thomas}}]{seipt2019ultrafast}%
  \BibitemOpen
  \bibfield  {author} {\bibinfo {author} {\bibfnamefont {D.}~\bibnamefont
  {Seipt}}, \bibinfo {author} {\bibfnamefont {D.}~\bibnamefont {Del~Sorbo}},
  \bibinfo {author} {\bibfnamefont {C.~P.}\ \bibnamefont {Ridgers}},\ and\
  \bibinfo {author} {\bibfnamefont {A.~G.}\ \bibnamefont {Thomas}},\ }\bibfield
   {title} {\bibinfo {title} {Ultrafast polarization of an electron beam in an
  intense bichromatic laser field},\ }\href@noop {} {\bibfield  {journal}
  {\bibinfo  {journal} {Physical Review A}\ }\textbf {\bibinfo {volume}
  {100}},\ \bibinfo {pages} {061402} (\bibinfo {year} {2019})}\BibitemShut
  {NoStop}%
\bibitem [{\citenamefont {Rakitzis}\ \emph {et~al.}(2003)\citenamefont
  {Rakitzis}, \citenamefont {Samartzis}, \citenamefont {Toomes}, \citenamefont
  {Kitsopoulos}, \citenamefont {Brown}, \citenamefont {Balint-Kurti},
  \citenamefont {Vasyutinskii},\ and\ \citenamefont
  {Beswick}}]{rakitzis2003spin}%
  \BibitemOpen
  \bibfield  {author} {\bibinfo {author} {\bibfnamefont {T.}~\bibnamefont
  {Rakitzis}}, \bibinfo {author} {\bibfnamefont {P.}~\bibnamefont {Samartzis}},
  \bibinfo {author} {\bibfnamefont {R.}~\bibnamefont {Toomes}}, \bibinfo
  {author} {\bibfnamefont {T.}~\bibnamefont {Kitsopoulos}}, \bibinfo {author}
  {\bibfnamefont {A.}~\bibnamefont {Brown}}, \bibinfo {author} {\bibfnamefont
  {G.}~\bibnamefont {Balint-Kurti}}, \bibinfo {author} {\bibfnamefont
  {O.}~\bibnamefont {Vasyutinskii}},\ and\ \bibinfo {author} {\bibfnamefont
  {J.}~\bibnamefont {Beswick}},\ }\bibfield  {title} {\bibinfo {title}
  {Spin-polarized hydrogen atoms from molecular photodissociation},\
  }\href@noop {} {\bibfield  {journal} {\bibinfo  {journal} {Science}\ }\textbf
  {\bibinfo {volume} {300}},\ \bibinfo {pages} {1936} (\bibinfo {year}
  {2003})}\BibitemShut {NoStop}%
\bibitem [{\citenamefont {Rakitzis}(2004)}]{rakitzis2004pulsed}%
  \BibitemOpen
  \bibfield  {author} {\bibinfo {author} {\bibfnamefont {T.~P.}\ \bibnamefont
  {Rakitzis}},\ }\bibfield  {title} {\bibinfo {title} {Pulsed-laser production
  and detection of spin-polarized hydrogen atoms},\ }\href@noop {} {\bibfield
  {journal} {\bibinfo  {journal} {ChemPhysChem}\ }\textbf {\bibinfo {volume}
  {5}},\ \bibinfo {pages} {1489} (\bibinfo {year} {2004})}\BibitemShut
  {NoStop}%
\bibitem [{\citenamefont {Sofikitis}\ \emph {et~al.}(2008)\citenamefont
  {Sofikitis}, \citenamefont {Rubio-Lago}, \citenamefont {Bougas},
  \citenamefont {Alexander},\ and\ \citenamefont
  {Rakitzis}}]{sofikitis2008laser}%
  \BibitemOpen
  \bibfield  {author} {\bibinfo {author} {\bibfnamefont {D.}~\bibnamefont
  {Sofikitis}}, \bibinfo {author} {\bibfnamefont {L.}~\bibnamefont
  {Rubio-Lago}}, \bibinfo {author} {\bibfnamefont {L.}~\bibnamefont {Bougas}},
  \bibinfo {author} {\bibfnamefont {A.~J.}\ \bibnamefont {Alexander}},\ and\
  \bibinfo {author} {\bibfnamefont {T.~P.}\ \bibnamefont {Rakitzis}},\
  }\bibfield  {title} {\bibinfo {title} {Laser detection of spin-polarized
  hydrogen from hcl and hbr photodissociation: Comparison of h-and halogen-atom
  polarizations},\ }\href@noop {} {\bibfield  {journal} {\bibinfo  {journal}
  {The Journal of chemical physics}\ }\textbf {\bibinfo {volume} {129}},\
  \bibinfo {pages} {144302} (\bibinfo {year} {2008})}\BibitemShut {NoStop}%
\bibitem [{\citenamefont {Sofikitis}\ \emph {et~al.}(2018)\citenamefont
  {Sofikitis}, \citenamefont {Kannis}, \citenamefont {Boulogiannis},\ and\
  \citenamefont {Rakitzis}}]{sofikitis2018ultrahigh}%
  \BibitemOpen
  \bibfield  {author} {\bibinfo {author} {\bibfnamefont {D.}~\bibnamefont
  {Sofikitis}}, \bibinfo {author} {\bibfnamefont {C.~S.}\ \bibnamefont
  {Kannis}}, \bibinfo {author} {\bibfnamefont {G.~K.}\ \bibnamefont
  {Boulogiannis}},\ and\ \bibinfo {author} {\bibfnamefont {T.~P.}\ \bibnamefont
  {Rakitzis}},\ }\bibfield  {title} {\bibinfo {title} {Ultrahigh-density
  spin-polarized h and d observed via magnetization quantum beats},\
  }\href@noop {} {\bibfield  {journal} {\bibinfo  {journal} {Physical review
  letters}\ }\textbf {\bibinfo {volume} {121}},\ \bibinfo {pages} {083001}
  (\bibinfo {year} {2018})}\BibitemShut {NoStop}%
\bibitem [{\citenamefont {Wen}\ \emph {et~al.}(2019)\citenamefont {Wen},
  \citenamefont {Tamburini},\ and\ \citenamefont {Keitel}}]{wen2019polarized}%
  \BibitemOpen
  \bibfield  {author} {\bibinfo {author} {\bibfnamefont {M.}~\bibnamefont
  {Wen}}, \bibinfo {author} {\bibfnamefont {M.}~\bibnamefont {Tamburini}},\
  and\ \bibinfo {author} {\bibfnamefont {C.~H.}\ \bibnamefont {Keitel}},\
  }\bibfield  {title} {\bibinfo {title} {Polarized laser-wakefield-accelerated
  kiloampere electron beams},\ }\href@noop {} {\bibfield  {journal} {\bibinfo
  {journal} {Physical review letters}\ }\textbf {\bibinfo {volume} {122}},\
  \bibinfo {pages} {214801} (\bibinfo {year} {2019})}\BibitemShut {NoStop}%
\bibitem [{\citenamefont {Wu}\ \emph {et~al.}(2019{\natexlab{a}})\citenamefont
  {Wu}, \citenamefont {Ji}, \citenamefont {Geng}, \citenamefont {Yu},
  \citenamefont {Wang}, \citenamefont {Feng}, \citenamefont {Guo},
  \citenamefont {Wang}, \citenamefont {Qin}, \citenamefont {Yan} \emph
  {et~al.}}]{wu2019polarized}%
  \BibitemOpen
  \bibfield  {author} {\bibinfo {author} {\bibfnamefont {Y.}~\bibnamefont
  {Wu}}, \bibinfo {author} {\bibfnamefont {L.}~\bibnamefont {Ji}}, \bibinfo
  {author} {\bibfnamefont {X.}~\bibnamefont {Geng}}, \bibinfo {author}
  {\bibfnamefont {Q.}~\bibnamefont {Yu}}, \bibinfo {author} {\bibfnamefont
  {N.}~\bibnamefont {Wang}}, \bibinfo {author} {\bibfnamefont {B.}~\bibnamefont
  {Feng}}, \bibinfo {author} {\bibfnamefont {Z.}~\bibnamefont {Guo}}, \bibinfo
  {author} {\bibfnamefont {W.}~\bibnamefont {Wang}}, \bibinfo {author}
  {\bibfnamefont {C.}~\bibnamefont {Qin}}, \bibinfo {author} {\bibfnamefont
  {X.}~\bibnamefont {Yan}}, \emph {et~al.},\ }\bibfield  {title} {\bibinfo
  {title} {Polarized electron-beam acceleration driven by vortex laser
  pulses},\ }\href@noop {} {\bibfield  {journal} {\bibinfo  {journal} {New
  Journal of Physics}\ }\textbf {\bibinfo {volume} {21}},\ \bibinfo {pages}
  {073052} (\bibinfo {year} {2019}{\natexlab{a}})}\BibitemShut {NoStop}%
\bibitem [{\citenamefont {Wu}\ \emph {et~al.}(2019{\natexlab{b}})\citenamefont
  {Wu}, \citenamefont {Ji}, \citenamefont {Geng}, \citenamefont {Yu},
  \citenamefont {Wang}, \citenamefont {Feng}, \citenamefont {Guo},
  \citenamefont {Wang}, \citenamefont {Qin}, \citenamefont {Yan} \emph
  {et~al.}}]{wu2019polarized1}%
  \BibitemOpen
  \bibfield  {author} {\bibinfo {author} {\bibfnamefont {Y.}~\bibnamefont
  {Wu}}, \bibinfo {author} {\bibfnamefont {L.}~\bibnamefont {Ji}}, \bibinfo
  {author} {\bibfnamefont {X.}~\bibnamefont {Geng}}, \bibinfo {author}
  {\bibfnamefont {Q.}~\bibnamefont {Yu}}, \bibinfo {author} {\bibfnamefont
  {N.}~\bibnamefont {Wang}}, \bibinfo {author} {\bibfnamefont {B.}~\bibnamefont
  {Feng}}, \bibinfo {author} {\bibfnamefont {Z.}~\bibnamefont {Guo}}, \bibinfo
  {author} {\bibfnamefont {W.}~\bibnamefont {Wang}}, \bibinfo {author}
  {\bibfnamefont {C.}~\bibnamefont {Qin}}, \bibinfo {author} {\bibfnamefont
  {X.}~\bibnamefont {Yan}}, \emph {et~al.},\ }\bibfield  {title} {\bibinfo
  {title} {Polarized electron acceleration in beam-driven plasma wakefield
  based on density down-ramp injection},\ }\href@noop {} {\bibfield  {journal}
  {\bibinfo  {journal} {Physical Review E}\ }\textbf {\bibinfo {volume}
  {100}},\ \bibinfo {pages} {043202} (\bibinfo {year}
  {2019}{\natexlab{b}})}\BibitemShut {NoStop}%
\bibitem [{\citenamefont {Wu}\ \emph {et~al.}(2020)\citenamefont {Wu},
  \citenamefont {Ji}, \citenamefont {Geng}, \citenamefont {Thomas},
  \citenamefont {B{\"u}scher}, \citenamefont {Pukhov}, \citenamefont
  {H{\"u}tzen}, \citenamefont {Zhang}, \citenamefont {Shen},\ and\
  \citenamefont {Li}}]{wu2020spin}%
  \BibitemOpen
  \bibfield  {author} {\bibinfo {author} {\bibfnamefont {Y.}~\bibnamefont
  {Wu}}, \bibinfo {author} {\bibfnamefont {L.}~\bibnamefont {Ji}}, \bibinfo
  {author} {\bibfnamefont {X.}~\bibnamefont {Geng}}, \bibinfo {author}
  {\bibfnamefont {J.}~\bibnamefont {Thomas}}, \bibinfo {author} {\bibfnamefont
  {M.}~\bibnamefont {B{\"u}scher}}, \bibinfo {author} {\bibfnamefont
  {A.}~\bibnamefont {Pukhov}}, \bibinfo {author} {\bibfnamefont
  {A.}~\bibnamefont {H{\"u}tzen}}, \bibinfo {author} {\bibfnamefont
  {L.}~\bibnamefont {Zhang}}, \bibinfo {author} {\bibfnamefont
  {B.}~\bibnamefont {Shen}},\ and\ \bibinfo {author} {\bibfnamefont
  {R.}~\bibnamefont {Li}},\ }\bibfield  {title} {\bibinfo {title} {Spin filter
  for polarized electron acceleration in plasma wakefields},\ }\href@noop {}
  {\bibfield  {journal} {\bibinfo  {journal} {Physical review applied}\
  }\textbf {\bibinfo {volume} {13}},\ \bibinfo {pages} {044064} (\bibinfo
  {year} {2020})}\BibitemShut {NoStop}%
\bibitem [{\citenamefont {Leemans}\ \emph {et~al.}(2014)\citenamefont
  {Leemans}, \citenamefont {Gonsalves}, \citenamefont {Mao}, \citenamefont
  {Nakamura}, \citenamefont {Benedetti}, \citenamefont {Schroeder},
  \citenamefont {T{\'o}th}, \citenamefont {Daniels}, \citenamefont
  {Mittelberger}, \citenamefont {Bulanov} \emph {et~al.}}]{leemans2014multi}%
  \BibitemOpen
  \bibfield  {author} {\bibinfo {author} {\bibfnamefont {W.}~\bibnamefont
  {Leemans}}, \bibinfo {author} {\bibfnamefont {A.}~\bibnamefont {Gonsalves}},
  \bibinfo {author} {\bibfnamefont {H.-S.}\ \bibnamefont {Mao}}, \bibinfo
  {author} {\bibfnamefont {K.}~\bibnamefont {Nakamura}}, \bibinfo {author}
  {\bibfnamefont {C.}~\bibnamefont {Benedetti}}, \bibinfo {author}
  {\bibfnamefont {C.}~\bibnamefont {Schroeder}}, \bibinfo {author}
  {\bibfnamefont {C.}~\bibnamefont {T{\'o}th}}, \bibinfo {author}
  {\bibfnamefont {J.}~\bibnamefont {Daniels}}, \bibinfo {author} {\bibfnamefont
  {D.}~\bibnamefont {Mittelberger}}, \bibinfo {author} {\bibfnamefont
  {S.}~\bibnamefont {Bulanov}}, \emph {et~al.},\ }\bibfield  {title} {\bibinfo
  {title} {Multi-gev electron beams from capillary-discharge-guided subpetawatt
  laser pulses in the self-trapping regime},\ }\href@noop {} {\bibfield
  {journal} {\bibinfo  {journal} {Physical review letters}\ }\textbf {\bibinfo
  {volume} {113}},\ \bibinfo {pages} {245002} (\bibinfo {year}
  {2014})}\BibitemShut {NoStop}%
\bibitem [{\citenamefont {Gonsalves}\ \emph {et~al.}(2019)\citenamefont
  {Gonsalves}, \citenamefont {Nakamura}, \citenamefont {Daniels}, \citenamefont
  {Benedetti}, \citenamefont {Pieronek}, \citenamefont {De~Raadt},
  \citenamefont {Steinke}, \citenamefont {Bin}, \citenamefont {Bulanov},
  \citenamefont {Van~Tilborg} \emph {et~al.}}]{gonsalves2019petawatt}%
  \BibitemOpen
  \bibfield  {author} {\bibinfo {author} {\bibfnamefont {A.}~\bibnamefont
  {Gonsalves}}, \bibinfo {author} {\bibfnamefont {K.}~\bibnamefont {Nakamura}},
  \bibinfo {author} {\bibfnamefont {J.}~\bibnamefont {Daniels}}, \bibinfo
  {author} {\bibfnamefont {C.}~\bibnamefont {Benedetti}}, \bibinfo {author}
  {\bibfnamefont {C.}~\bibnamefont {Pieronek}}, \bibinfo {author}
  {\bibfnamefont {T.}~\bibnamefont {De~Raadt}}, \bibinfo {author}
  {\bibfnamefont {S.}~\bibnamefont {Steinke}}, \bibinfo {author} {\bibfnamefont
  {J.}~\bibnamefont {Bin}}, \bibinfo {author} {\bibfnamefont {S.}~\bibnamefont
  {Bulanov}}, \bibinfo {author} {\bibfnamefont {J.}~\bibnamefont
  {Van~Tilborg}}, \emph {et~al.},\ }\bibfield  {title} {\bibinfo {title}
  {Petawatt laser guiding and electron beam acceleration to 8 gev in a
  laser-heated capillary discharge waveguide},\ }\href@noop {} {\bibfield
  {journal} {\bibinfo  {journal} {Physical review letters}\ }\textbf {\bibinfo
  {volume} {122}},\ \bibinfo {pages} {084801} (\bibinfo {year}
  {2019})}\BibitemShut {NoStop}%
\bibitem [{\citenamefont {Gol’dman}(1964)}]{gol1964intensity}%
  \BibitemOpen
  \bibfield  {author} {\bibinfo {author} {\bibfnamefont {I.}~\bibnamefont
  {Gol’dman}},\ }\bibfield  {title} {\bibinfo {title} {Intensity effects in
  compton scattering},\ }\href@noop {} {\bibfield  {journal} {\bibinfo
  {journal} {Sov. Phys. JETP}\ }\textbf {\bibinfo {volume} {19}},\ \bibinfo
  {pages} {954} (\bibinfo {year} {1964})}\BibitemShut {NoStop}%
\bibitem [{\citenamefont {Nikishov}\ and\ \citenamefont
  {Ritus}(1964)}]{nikishov1964quantum}%
  \BibitemOpen
  \bibfield  {author} {\bibinfo {author} {\bibfnamefont {A.}~\bibnamefont
  {Nikishov}}\ and\ \bibinfo {author} {\bibfnamefont {V.}~\bibnamefont
  {Ritus}},\ }\bibfield  {title} {\bibinfo {title} {Quantum processes in the
  field of a plane electromagnetic wave and in a constant field. i},\
  }\href@noop {} {\bibfield  {journal} {\bibinfo  {journal} {Sov. Phys. JETP}\
  }\textbf {\bibinfo {volume} {19}},\ \bibinfo {pages} {529} (\bibinfo {year}
  {1964})}\BibitemShut {NoStop}%
\bibitem [{\citenamefont {Ritus}(1985)}]{ritus1985quantum}%
  \BibitemOpen
  \bibfield  {author} {\bibinfo {author} {\bibfnamefont {V.}~\bibnamefont
  {Ritus}},\ }\bibfield  {title} {\bibinfo {title} {Quantum effects of the
  interaction of elementary particles with an intense electromagnetic field},\
  }\href@noop {} {\bibfield  {journal} {\bibinfo  {journal} {J. Sov. Laser
  Res.;(United States)}\ }\textbf {\bibinfo {volume} {6}} (\bibinfo {year}
  {1985})}\BibitemShut {NoStop}%
\bibitem [{\citenamefont {Gong}\ \emph {et~al.}(2021)\citenamefont {Gong},
  \citenamefont {Hatsagortsyan},\ and\ \citenamefont
  {Keitel}}]{gong2021retrieving}%
  \BibitemOpen
  \bibfield  {author} {\bibinfo {author} {\bibfnamefont {Z.}~\bibnamefont
  {Gong}}, \bibinfo {author} {\bibfnamefont {K.~Z.}\ \bibnamefont
  {Hatsagortsyan}},\ and\ \bibinfo {author} {\bibfnamefont {C.~H.}\
  \bibnamefont {Keitel}},\ }\bibfield  {title} {\bibinfo {title} {Retrieving
  transient magnetic fields of ultrarelativistic laser plasma via ejected
  electron polarization},\ }\href@noop {} {\bibfield  {journal} {\bibinfo
  {journal} {Physical review letters}\ }\textbf {\bibinfo {volume} {127}},\
  \bibinfo {pages} {165002} (\bibinfo {year} {2021})}\BibitemShut {NoStop}%
\bibitem [{\citenamefont {Gong}\ \emph {et~al.}(2019)\citenamefont {Gong},
  \citenamefont {Hu}, \citenamefont {Yu}, \citenamefont {Shou}, \citenamefont
  {Arefiev},\ and\ \citenamefont {Yan}}]{gong2019radiation}%
  \BibitemOpen
  \bibfield  {author} {\bibinfo {author} {\bibfnamefont {Z.}~\bibnamefont
  {Gong}}, \bibinfo {author} {\bibfnamefont {R.}~\bibnamefont {Hu}}, \bibinfo
  {author} {\bibfnamefont {J.}~\bibnamefont {Yu}}, \bibinfo {author}
  {\bibfnamefont {Y.}~\bibnamefont {Shou}}, \bibinfo {author} {\bibfnamefont
  {A.}~\bibnamefont {Arefiev}},\ and\ \bibinfo {author} {\bibfnamefont
  {X.}~\bibnamefont {Yan}},\ }\bibfield  {title} {\bibinfo {title} {Radiation
  rebound and quantum splash in electron-laser collisions},\ }\href@noop {}
  {\bibfield  {journal} {\bibinfo  {journal} {Physical Review Accelerators and
  Beams}\ }\textbf {\bibinfo {volume} {22}},\ \bibinfo {pages} {093401}
  (\bibinfo {year} {2019})}\BibitemShut {NoStop}%
\bibitem [{\citenamefont {Tang}\ \emph {et~al.}(2021)\citenamefont {Tang},
  \citenamefont {Gong}, \citenamefont {Yu}, \citenamefont {Shou},\ and\
  \citenamefont {Yan}}]{tang2021radiative}%
  \BibitemOpen
  \bibfield  {author} {\bibinfo {author} {\bibfnamefont {Y.}~\bibnamefont
  {Tang}}, \bibinfo {author} {\bibfnamefont {Z.}~\bibnamefont {Gong}}, \bibinfo
  {author} {\bibfnamefont {J.}~\bibnamefont {Yu}}, \bibinfo {author}
  {\bibfnamefont {Y.}~\bibnamefont {Shou}},\ and\ \bibinfo {author}
  {\bibfnamefont {X.}~\bibnamefont {Yan}},\ }\bibfield  {title} {\bibinfo
  {title} {Radiative polarization dynamics of relativistic electrons in an
  intense electromagnetic field},\ }\href@noop {} {\bibfield  {journal}
  {\bibinfo  {journal} {Physical Review A}\ }\textbf {\bibinfo {volume}
  {103}},\ \bibinfo {pages} {042807} (\bibinfo {year} {2021})}\BibitemShut
  {NoStop}%
\bibitem [{\citenamefont {Thomas}(1926)}]{thomas1926motion}%
  \BibitemOpen
  \bibfield  {author} {\bibinfo {author} {\bibfnamefont {L.~H.}\ \bibnamefont
  {Thomas}},\ }\bibfield  {title} {\bibinfo {title} {The motion of the spinning
  electron},\ }\href@noop {} {\bibfield  {journal} {\bibinfo  {journal}
  {Nature}\ }\textbf {\bibinfo {volume} {117}},\ \bibinfo {pages} {514}
  (\bibinfo {year} {1926})}\BibitemShut {NoStop}%
\bibitem [{\citenamefont {Thomas}(1927)}]{thomas1927kinematics}%
  \BibitemOpen
  \bibfield  {author} {\bibinfo {author} {\bibfnamefont {L.~H.}\ \bibnamefont
  {Thomas}},\ }\bibfield  {title} {\bibinfo {title} {I. the kinematics of an
  electron with an axis},\ }\href@noop {} {\bibfield  {journal} {\bibinfo
  {journal} {The London, Edinburgh, and Dublin Philosophical Magazine and
  Journal of Science}\ }\textbf {\bibinfo {volume} {3}},\ \bibinfo {pages} {1}
  (\bibinfo {year} {1927})}\BibitemShut {NoStop}%
\bibitem [{\citenamefont {Bargmann}\ \emph {et~al.}(1959)\citenamefont
  {Bargmann}, \citenamefont {Michel},\ and\ \citenamefont
  {Telegdi}}]{bargmann1959precession}%
  \BibitemOpen
  \bibfield  {author} {\bibinfo {author} {\bibfnamefont {V.}~\bibnamefont
  {Bargmann}}, \bibinfo {author} {\bibfnamefont {L.}~\bibnamefont {Michel}},\
  and\ \bibinfo {author} {\bibfnamefont {V.}~\bibnamefont {Telegdi}},\
  }\bibfield  {title} {\bibinfo {title} {Precession of the polarization of
  particles moving in a homogeneous electromagnetic field},\ }\href@noop {}
  {\bibfield  {journal} {\bibinfo  {journal} {Physical Review Letters}\
  }\textbf {\bibinfo {volume} {2}},\ \bibinfo {pages} {435} (\bibinfo {year}
  {1959})}\BibitemShut {NoStop}%
\bibitem [{\citenamefont {Duclous}\ \emph {et~al.}(2010)\citenamefont
  {Duclous}, \citenamefont {Kirk},\ and\ \citenamefont
  {Bell}}]{duclous2010monte}%
  \BibitemOpen
  \bibfield  {author} {\bibinfo {author} {\bibfnamefont {R.}~\bibnamefont
  {Duclous}}, \bibinfo {author} {\bibfnamefont {J.~G.}\ \bibnamefont {Kirk}},\
  and\ \bibinfo {author} {\bibfnamefont {A.~R.}\ \bibnamefont {Bell}},\
  }\bibfield  {title} {\bibinfo {title} {Monte carlo calculations of pair
  production in high-intensity laser--plasma interactions},\ }\href@noop {}
  {\bibfield  {journal} {\bibinfo  {journal} {Plasma Physics and Controlled
  Fusion}\ }\textbf {\bibinfo {volume} {53}},\ \bibinfo {pages} {015009}
  (\bibinfo {year} {2010})}\BibitemShut {NoStop}%
\bibitem [{\citenamefont {Ba{\u{\i}}er}(1972)}]{bauier1972radiative}%
  \BibitemOpen
  \bibfield  {author} {\bibinfo {author} {\bibfnamefont {V.}~\bibnamefont
  {Ba{\u{\i}}er}},\ }\bibfield  {title} {\bibinfo {title} {Radiative
  polarization of electrons in storage rings},\ }\href@noop {} {\bibfield
  {journal} {\bibinfo  {journal} {Soviet Physics Uspekhi}\ }\textbf {\bibinfo
  {volume} {14}},\ \bibinfo {pages} {695} (\bibinfo {year} {1972})}\BibitemShut
  {NoStop}%
\bibitem [{\citenamefont {Guo}\ \emph {et~al.}(2020)\citenamefont {Guo},
  \citenamefont {Wang}, \citenamefont {Shaisultanov}, \citenamefont {Wan},
  \citenamefont {Xu}, \citenamefont {Chen}, \citenamefont {Hatsagortsyan},\
  and\ \citenamefont {Li}}]{guo2020stochasticity}%
  \BibitemOpen
  \bibfield  {author} {\bibinfo {author} {\bibfnamefont {R.-T.}\ \bibnamefont
  {Guo}}, \bibinfo {author} {\bibfnamefont {Y.}~\bibnamefont {Wang}}, \bibinfo
  {author} {\bibfnamefont {R.}~\bibnamefont {Shaisultanov}}, \bibinfo {author}
  {\bibfnamefont {F.}~\bibnamefont {Wan}}, \bibinfo {author} {\bibfnamefont
  {Z.-F.}\ \bibnamefont {Xu}}, \bibinfo {author} {\bibfnamefont {Y.-Y.}\
  \bibnamefont {Chen}}, \bibinfo {author} {\bibfnamefont {K.~Z.}\ \bibnamefont
  {Hatsagortsyan}},\ and\ \bibinfo {author} {\bibfnamefont {J.-X.}\
  \bibnamefont {Li}},\ }\bibfield  {title} {\bibinfo {title} {Stochasticity in
  radiative polarization of ultrarelativistic electrons in an ultrastrong laser
  pulse},\ }\href {https://doi.org/10.1103/PhysRevResearch.2.033483} {\bibfield
   {journal} {\bibinfo  {journal} {Phys. Rev. Research}\ }\textbf {\bibinfo
  {volume} {2}},\ \bibinfo {pages} {033483} (\bibinfo {year}
  {2020})}\BibitemShut {NoStop}%
\bibitem [{\citenamefont {Li}\ \emph {et~al.}(2020)\citenamefont {Li},
  \citenamefont {Chen}, \citenamefont {Wang},\ and\ \citenamefont
  {Hu}}]{li2020production}%
  \BibitemOpen
  \bibfield  {author} {\bibinfo {author} {\bibfnamefont {Y.-F.}\ \bibnamefont
  {Li}}, \bibinfo {author} {\bibfnamefont {Y.-Y.}\ \bibnamefont {Chen}},
  \bibinfo {author} {\bibfnamefont {W.-M.}\ \bibnamefont {Wang}},\ and\
  \bibinfo {author} {\bibfnamefont {H.-S.}\ \bibnamefont {Hu}},\ }\bibfield
  {title} {\bibinfo {title} {Production of highly polarized positron beams via
  helicity transfer from polarized electrons in a strong laser field},\ }\href
  {https://doi.org/10.1103/PhysRevLett.125.044802} {\bibfield  {journal}
  {\bibinfo  {journal} {Phys. Rev. Lett.}\ }\textbf {\bibinfo {volume} {125}},\
  \bibinfo {pages} {044802} (\bibinfo {year} {2020})}\BibitemShut {NoStop}%
\bibitem [{\citenamefont {Gibbon}(2005)}]{gibbon2005short}%
  \BibitemOpen
  \bibfield  {author} {\bibinfo {author} {\bibfnamefont {P.}~\bibnamefont
  {Gibbon}},\ }\href {https://doi.org/10.1142/p116} {\emph {\bibinfo {title}
  {Short Pulse Laser Interactions with Matter}}}\ (\bibinfo  {publisher}
  {Imperial College Press},\ \bibinfo {address} {London},\ \bibinfo {year}
  {2005})\BibitemShut {NoStop}%
\end{thebibliography}%

\end{document}